\documentclass[12pt,preprint]{aastex}

\usepackage{amsmath}

\newcommand\cm{{\rm cm}}
\newcommand\s{{\rm s}}
\newcommand\g{{\rm g}}
\newcommand\erg{{\rm erg}}
\newcommand\K{{\rm K}}
\newcommand\yr{{\rm yr}}
\newcommand\Myr{{\rm Myr}}

\newcommand\km{{\rm km}}
\newcommand\kms{{\rm km\, s^{-1}}}
\newcommand\pc{{\rm\,pc}}

\newcommand\simgt{\lower.5ex\hbox{$\; \buildrel > \over \sim \;$}}
\newcommand\simlt{\lower.5ex\hbox{$\; \buildrel < \over \sim \;$}}

\slugcomment{Accepted for Publication in ApJ}

\shorttitle{Thermal and Magnetorotational Instabilities}
\shortauthors{Piontek \& Ostriker}

\begin{document}


\title{Thermal and Magnetorotational Instability in the ISM: 
Two-Dimensional Numerical Simulations}


\author{Robert A. Piontek and Eve C. Ostriker}
\affil{Department of Astronomy\\University of Maryland\\
    College Park, MD  20742-2421}
\email{rpiontek@astro.umd.edu, ostriker@astro.umd.edu}



\begin{abstract}  
  The structure and dynamics of diffuse gas in the Milky Way and other
  disk galaxies may be strongly influenced by thermal and
  magnetorotational instabilities (TI and MRI) on scales $\sim 1-100$
  pc.  We initiate a study of these processes, using two-dimensional
  numerical hydrodynamic and magnetohydrodynamic (MHD) simulations
  with conditions appropriate for the atomic interstellar medium
  (ISM).  Our simulations incorporate thermal conduction, and adopt
  local ``shearing-periodic'' equations of motion and boundary
  conditions to study dynamics of a $(100\ \pc)^2$ radial-vertical
  section of the disk.  We demonstrate, consistent with previous work,
  that nonlinear development of ``pure TI'' produces a network of
  filaments that condense into cold clouds at their intersections,
  yielding a distinct two-phase warm/cold medium within $\sim 20$ Myr.
  TI-driven turbulent motions of the clouds and warm intercloud medium
  are present, but saturate at quite subsonic amplitudes for uniform
  initial $P/k=2000 \ \K \ \cm^{-3}$.  MRI has previously been studied
  in near-uniform media; our simulations include both TI+MRI models,
  which begin from uniform-density conditions, and cloud+MRI models,
  which begin with a two-phase cloudy medium.  Both the TI+MRI and
  cloud+MRI models show that MRI develops within a few galactic
  orbital times, just as for a uniform medium.  The mean separation
  between clouds can affect which MRI mode dominates the evolution.
  Provided intercloud separations do not exceed half the MRI
  wavelength, we find the MRI growth rates are similar to those for
  the corresponding uniform medium.  This opens the possibility, if
  low cloud volume filling factors increase MRI dissipation times
  compared to those in a uniform medium, that MRI-driven motions in
  the ISM could reach amplitudes comparable to observed HI turbulent
  linewidths.
\end{abstract}




\section{Introduction}

The Galactic interstellar medium (ISM) is characterized by complex
spatial distributions of density, temperature, and magnetic fields, as
well as a turbulent velocity field that animates the whole system.
The relative proportions of ISM gas in different thermal/ionization
phases, and their respective dynamical states, may reflect many
contributing physical processes of varying importance throughout the
Milky Way (or external galaxies).  Even considering just the Galaxy's
atomic gas component, observable in HI emission and absorption, a wide
variety of temperatures and pervasive high-amplitude turbulence is
inferred \citep{hei03}, and a number of different physical processes
may collude or compete in establishing these conditions.

In the traditional picture of the ISM, turbulence in atomic gas is
primarily attributed to the lingering effects of supernova blast waves
that sweep through the ISM \citep{cox74, mck77, spi78}.  Densities and
temperatures of atomic gas are expected to lie preferentially near
either the warm or cold stable thermal equilibria available given
heating primarily by the photoelectric effect on small grains
\citep{wol95,wol03}.  Thermal instability (TI) is believed to play an
important role in maintaining gas near the stable equilibria
\citep{fie65}.

Certain potential difficulties with this picture motivate an effort to
explore effects not emphasized in the traditional model.  In
particular, because energetic stellar inputs are intermittent in space
and time, while turbulence is directly or indirectly inferred to
pervade the whole atomic ISM, it is valuable to assess alternative
spatially/temporally {\it distributed} turbulent driving mechanisms.
Candidate mechanisms recently proposed for driving turbulence include
both TI \citep{koy02, kri02a, kri02b} and the magnetorotational
instability (MRI) \citep{sel99,kim03}.  In addition to uncertainties
about the source of turbulence in HI gas, other puzzles surrounding HI
temperatures (e.g. \citet{kal85,ver94,spi95,fit97}) have grown more pressing
with recent observations \citep{hei01,hei03}.  Namely, the Heiles and
Troland observations suggest that significant HI gas ($\simgt 48\%$)
could be in the thermally-unstable temperature regime between 500-5000
K. Using observational evidence from various tracers, \citet{jen03}
has also recently argued that very large pressures and other large departures
from dynamical and thermal equilibrium are common in the ISM, and
indicate rapid changes likely driven by turbulence.  To assess and
interpret this evidence theoretically, it is necessary to understand
the nonlinear development of TI, the effects of independent dynamical
ISM processes on TI, and the ability in general of
magnetohydrodyanmic (MHD) turbulence to heat and cool ISM gas via
shocks, compressions, and rarefactions.

In recent years, direct numerical simulation has become an
increasingly important tool in theoretical investigation of the ISM's
structure and dynamics, and has played a key role in promoting the
increasingly popular notion of the ISM as a ``phase continuum''.  In
MHD (or hydrodynamic) simulations, the evolution of gas in the
computational domain is formalized in terms of time-dependent flow
equations with appropriate source terms to describe externally-imposed
effects.  Fully realistic computational ISM models will ultimately
require numerical simulations with a comprehensive array of physics
inputs.  Recent work towards this goal that address
turbulent driving and temperature/density probability
distribution functions (PDFs) include the three-dimensional (3D)
simulations of \citet{kor99}, \citet{dea00}, \citet{wad01b}, and
\citet{mac01}; and the two-dimensional (2D) simulations of
\citet{ros95}, \citet{wad00}, \citet{wad01a}, and \citet{gaz01}.
Among other physics inputs, all of these simulations include modeled
effects of star formation, with either supernova-like or stellar-like
localized heating events that lead to expanding flows.  For some of 
these models, the cooling functions also permit TI in certain density
regimes.

Since many of the individual processes affecting the ISM's structure
and dynamics are not well understood, in addition to comprehensive
physical modeling, it is also valuable to perform numerical
simulations that focus more narrowly on a single process, or on a few
processes that potentially may interact strongly.  This controlled
approach can yield significant insight into the relative importance of
multiple effects in complex systems such as the ISM.  Using models
that omit supernova and stellar energy inputs, it is possible to sort
out, for example, whether the appearance of phase continua in
density/temperature PDFs requires localized thermal energy inputs, or
can develop simply from the disruption of TI by moderate-amplitude
turbulence such as that driven by MRI.

Recent simulations that have focused on the nonlinear development of
TI under ISM conditions include \citet{hen99}, \citet{bur00},
\citet{vaz00}, \citet{san02}, \citet{kri02a,kri02b},
\citet{vaz03}. Previous simulations of MRI in 2D and 3D have focused
primarily on the situation in which the density is relatively uniform,
for application to accretion disks (e.g. \citet{haw92},
\citet{haw95a}, \citet{sto96}).  In recent work, \citet{kim03} began
study of MRI in the galactic context using isothermal simulations,
focusing on dense cloud formation due to the action of self-gravity
on turbulently-compressed regions.

In this work, we initiate a computational study aimed at understanding
how density, temperature, velocity, and magnetic field distributions
would develop in the diffuse ISM in the absence of localized stellar
energy input.  Of particular interest is the interaction between TI
and MRI.  TI tends to produce a cloudy medium, and this cloudy medium
may affect both the growth rate of MRI and its dissipation rate, and
hence the saturated-state turbulent amplitude that is determined by
balancing these rates.  On the other hand, the turbulence produced by
MRI may suppress and/or enhance TI by disrupting and/or initiating the
growth of dense condensations.  Evaluation of quasi-steady-state
properties such as the mean turbulent velocity amplitude and the distribution
of temperatures will await 3D simulations.  In the present work, which
employs 2D simulations, we focus on evaluation of our code's
performance for studies of thermally bistable media, and on analysis
of nonlinear development in models of pure TI, TI together with MRI,
and MRI in a medium of pre-existing clouds.

In \S 2, we describe our numerical methods and code tests.  In \S 3,
we present results from simulations of thermally unstable gas without
magnetic fields, and in \S 4 we present results of models in which
magnetic fields and sheared rotation have been added so that MRI
occurs.  Finally, in \S 5, we summarize our results, discuss their
implications, and make comparisons to previous work.

\section{Numerical Methods}
\label{numerics}
\subsection{Model Equations and Computational Algorithms}
  
We integrate the time-dependent equations of magnetohydrodynamics
using a version of the ZEUS-2D code \citep{sto92a,sto92b}.  ZEUS uses
a time-explicit, operator-split, finite difference method for solving
the MHD equations on a staggered mesh, capturing shocks via an
artificial viscosity.  Velocities and magnetic field vectors are
face-centered, while energy and mass density are volume-centered.
ZEUS employs the CT and MOC algorithms \citep{eva88, haw95b} to
maintain $\nabla \cdot {\bf B} = 0$ and ensure accurate propagation
of Alfv\'{e}n waves.
     
For the present study, we have implemented volumetric heating and
cooling terms, and a thermal conduction term.  We also model the
differential rotation of the background flow and the variation of the
stellar/dark matter gravitational potential in the local limit with
$x\equiv R-R_0 \ll R_0$, where $R_0$ is the galactocentric radius of
the center of our computational domain.  The equations we solve are
therefore:

\begin{equation}
\frac{\partial\rho}{\partial t}+ \boldsymbol{\nabla} \cdot (\rho
{\bf v}) = 0
\end{equation}

\begin{equation}
\frac{\partial{\bf v}}{\partial t}+
{\bf v}\cdot\boldsymbol{\nabla}{\bf v}=-\frac{\boldsymbol\nabla P}{\rho} + 
\frac{1}{4\pi\rho}(\boldsymbol{\nabla} \times {\bf B})
\times {\bf B}+ 2 q \Omega^{2}x\hat{x}-2\boldsymbol{\Omega}\times 
{\bf v}
\label{mom}
\end{equation}

\begin{equation}
\frac{\partial \mathcal{E} } { \partial t }
+{ \bf v}\cdot\boldsymbol{\nabla}\mathcal{E} = 
-(\mathcal{E} + P)\boldsymbol{\nabla}\cdot{\bf v}-\rho\mathcal{L}+
\boldsymbol{\nabla}\cdot(\mathcal{K}\boldsymbol{\nabla}T)
\label{energy}
\end{equation}

\begin{equation}
\frac{\partial {\bf B}}{\partial t}=\boldsymbol{\nabla \times}({\bf v}
\times {\bf B})
\end{equation}

All symbols have their usual meanings.  The net cooling per unit mass
is given by $\mathcal{L}=\rho\Lambda(\rho,T)-\Gamma$.  We adopt the
simple atomic ISM heating and cooling prescriptions of \citet{san02},
in which the cooling function, $\Lambda(\rho,T)$, is a piecewise
power-law fit to the detailed models of \citet{wol95}.  The heating
rate, $\Gamma$, is taken to be constant at 0.015 $\rm{erg\ s^{-1}
  g^{-1}}$.  In the tidal potential term of equation (\ref{mom}),
$q\equiv-d \ln \Omega / d \ln R$ is the local dimensionless shear
parameter, equal to unity for a flat rotation curve in which the
angular velocity $\Omega \propto R^{-1}$.

The present set of simulations is 2D, with the computational domain
representing a square sector in the radial-vertical ($x-z$) plane.  In
the local frame, the azimuthal direction $\hat \phi$ becomes the
$\hat{y}$ coordinate axis; $y$-velocities and magnetic field
components are present in our models, but $\frac{\partial}{\partial
  y}=0$ for all variables.  To reduce diffusion from advection in the
presence of background shear, we apply the velocity decomposition
method of \citet{kim01}.  We employ periodic boundary conditions in
the $\hat{z}$-direction, and shearing-periodic boundary conditions in
the $\hat{x}$-direction \citep{haw92,haw95a}.  This framework allows
us to incorporate realistic galactic shear, while avoiding numerical
artifacts associated with simpler boundary conditions.

Because cooling times can be very short, the energy equation update
from the net cooling terms is solved implicitly using Newton-Raphson
iteration.  At the start of each iteration the time step is initially
computed from the CFL condition using the sound speed, Alfv\'{e}n
speed, and conduction parameter.  This is followed by a call to the
cooling subroutine.  The change in temperature within each zone is
limited to ten percent of its initial value.  If this requirement is
not met for all cells in the grid, the time step is reduced by a
factor of two, and the implicit energy update is recalculated.  Tests
with our cooling function show that this time step restriction could
in principle become quite prohibitive if zones were far from thermal
equilibrium.  In practice, though, for our model simulations this is
typically not the case, and the time step is reduced once or twice at
most.

The update from the conduction operator is solved explicitly, using a
simple five point stencil for the spatial second derivative of
temperature (cf. \citet{pre92} equation 19.2.4).  In two dimensions
the CFL condition is $\Delta t < (\Delta x)^2 [nk/(\gamma-1)]/ (4\mathcal{K})$. As
\cite{koy03} have recently pointed out, the importance of
incorporating conduction in simulations which contain thermally
unstable gas has been occasionally overlooked in past work.  Without
conduction, the growth rates for thermal instability are largest at
the smallest scales, and unresolved growth at the grid scale may
occur.\footnote{Similar numerical difficulties arise if the Jeans
scale is not resolved in simulations of self-gravitating clouds
\citep{tru97}.} The inclusion of conduction, however, has a
stabilizing effect on TI at small scales, and the conduction parameter
can be adjusted to allow spatial resolution of TI on the computational
grid.  Here, we treat $\mathcal{K}$ as a parameter that may be freely
specified for numerical efficacy; we discuss the physical level of
conduction in the ISM below.

\subsection{Code Tests}

\begin{figure}
\plotone{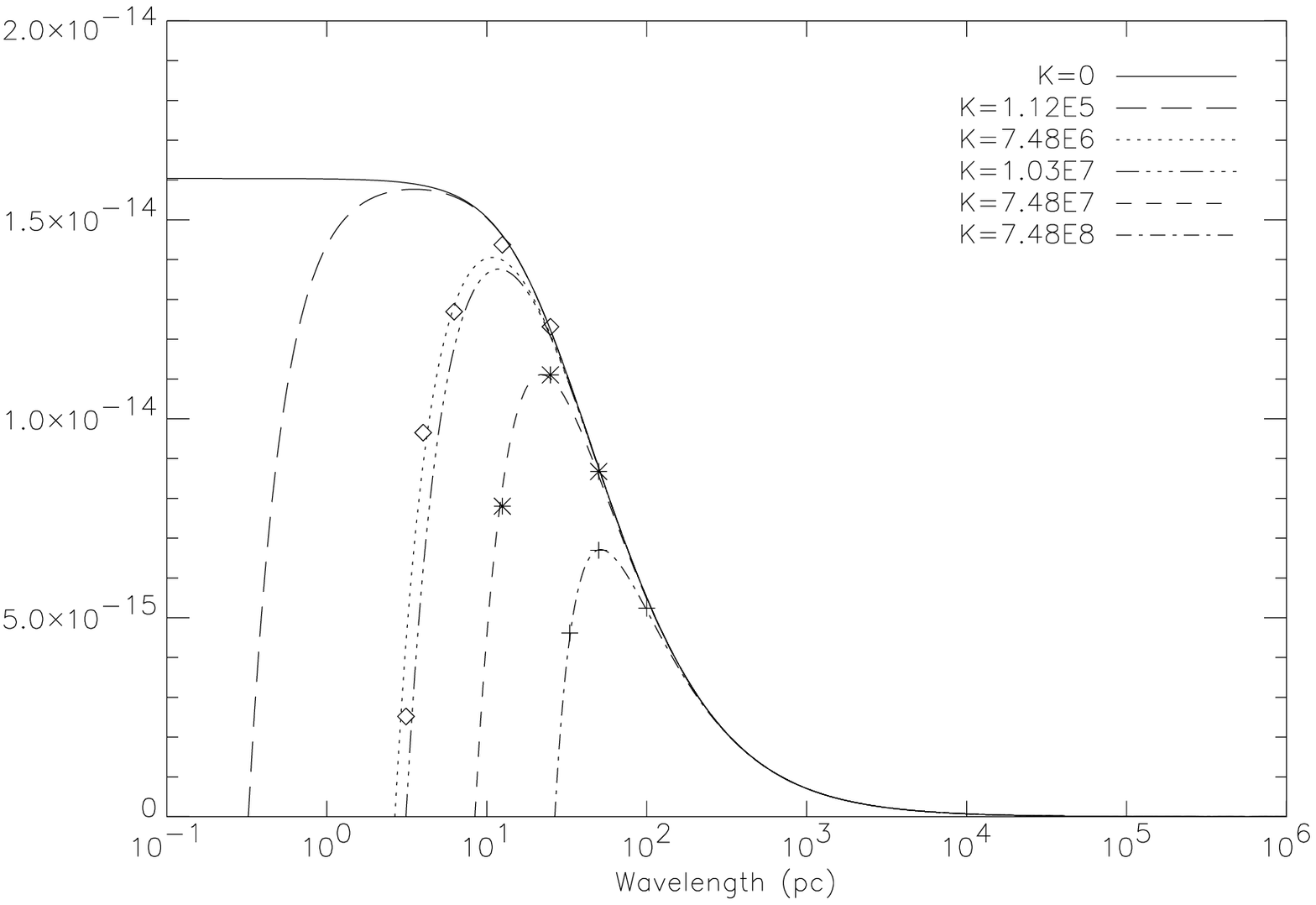}

\caption{
  Theoretical thermal instability growth rates \citep{fie65} for
  varying levels of conduction $\mathcal{K}=7.48 \times
  10^6,10^7,10^8$ (curves as indicated), overlayed with measured
  growth rates (points) from test simulations.  For reference we
  include the theoretical curves for $\mathcal{K}=0$ and for our
  chosen value of $\mathcal{K}=1.03 \times 10^7 \ \rm{erg \ cm^{-1} \ 
    K^{-1} \ s^{-1}}$, as well as using an estimate of the physical
  conduction in the ISM at 2000K, $\mathcal{K}=1.12 \times 10^5 \ 
  \rm{erg \ cm^{-1} \ K^{-1} \ s^{-1}}$. The asymptotic growth rate at
  small scales is $(2 \ \rm{Myr})^{-1}$ for the adopted
  parameters.
\label{fig1}}
\end{figure}

The ZEUS MHD code has undergone extensive numerical testing and has
been used in a wide variety of astrophysical investigations.  In
addition, we have tested the code without cooling and conduction and
have found it can accurately reproduce the linear growth rates of the
MRI for an adiabatic medium (see also \cite{haw92}). To test our
implementation of the heating, cooling, and conduction terms, we
performed 1D simulations to compare with the linear growth rates of
the thermal instability \citep{fie65}.  The models were initialized
with eigenmodes of the instability, and three levels of conduction
were chosen: $\mathcal{K}\in (7.48\times 10^6,7.48 \times 10^7,
7.48\times 10^8) \ \rm{erg \ cm^{-1} \ K^{-1} \ s^{-1}}$.  For these
tests, the grid was 128 zones and the box size $L=100$ pc.  The
initial density and pressure were set to $n=0.79 \ \cm^{-3}$ and
$P/k=2000 \ \cm^{-3} \ \K$, implying corresponding cutoffs for thermal
instability (``Field Length''), $\lambda_{\rm{F}}\in\{2.7,8.4,27\}$ pc
for our adopted cooling function (see \S 3.1)
\footnote{The Field length is $\lambda_{F} = 2\pi
\left[\frac{\rho^{2}\Lambda}{\mathcal{K}T}\left(1-\frac{\partial ln
\Lambda}{\partial ln T}\right)\right]^{-1/2}$ when
$\Lambda =$ function of $T$.}.   
In Figure~\ref{fig1} we plot the growth rates from the simulations on
top of the analytic curves.  The numerical growth rates are obtained by
measuring the logarithmic rate of change of the maximum density.  The
agreement between the analytic and numerical growth rates is quite
good. This test confirms that the newly added cooling and conduction
subroutines are working correctly, and is critical in assessing the
performance of the code as applied to multi-phase ISM simulations.
Note that at small scales TI is essentially isobaric, so that this
test demonstrates the ability of the code to maintain near-uniform
pressure via hydrodynamic flow to compensate for changes in
temperature driven by the cooling term (see eq. 3).

By comparison with simulations in which we set ${\cal K}=0$, tests
with non-zero ${\cal K}$ also confirm that conduction provides a
needed numerical stabilization.  When conduction is omitted, growth
rates for TI will always be greatest at the largest available
wavenumbers.  Our 2D tests with ${\cal K}=0$ have confirmed this is
indeed the case: simulations in which TI is seeded from random
perturbations form high density clouds which are the size of a single
grid zone.  Further 2D tests show that provided $\lambda_{\rm F}$ is
resolved by at least 8 zones, this grid-scale growth is suppressed.
For the models we shall present, the conduction parameter and grid
resolution were chosen such that we can adequately resolve all modes
for which TI is unstable.

Because we are modeling a medium containing very large density
contrasts, it is desirable to assess the evolution of contact
discontinuities between the high and low density regions, representing
cold and warm phases in pressure equilibrium. The diffusive smearing
of contact discontinuities is an inherent limitation of all finite
difference codes, but the numerical problem can be magnified with the
inclusion of a thermally bistable net cooling function.  As these
contact discontinuities are advected through the grid, upstream and
downstream zones adjacent to the discontinuity are set to intermediate
densities which may be thermally unstable.  Thermally unstable gas
adjoining the initial contact rapidly heats or cools to
reach a pressure different from the initial equilibrium, and this
can potentially introduce additional dynamics to the problem.

\begin{figure}
\epsscale{1.0}
\plotone{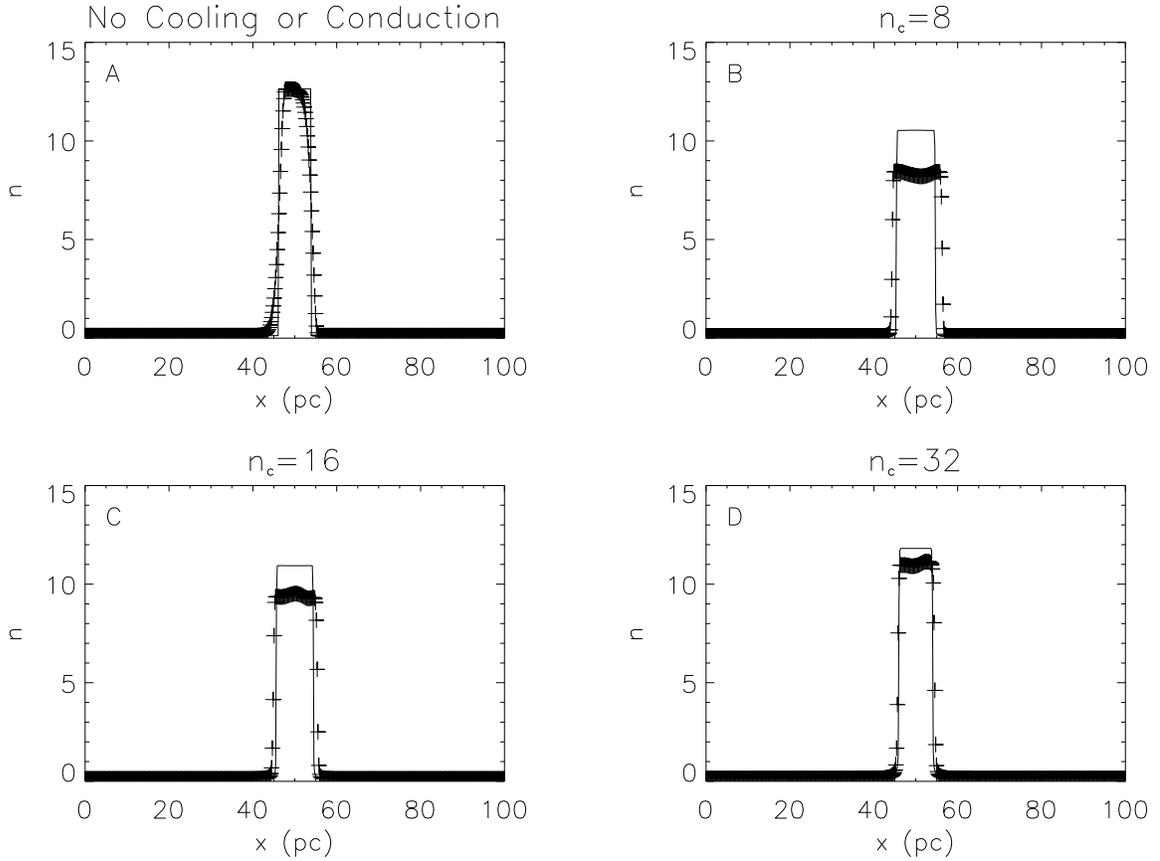}
\caption{Advection test results (crosses) compared with the initial profile (solid line) of a 1D cloud.  In Panel A we show results from the original ZEUS-2D code, without cooling and conduction.  In Panels B, C, and D we set the conduction parameter so that $n_c =$ 8, 16, and 32, respectively.  Advection test results improve as $n_c$ increases. 
\label{fig2}}
\end{figure}

To explore this numerical issue for the problem at hand, we have
performed 1D advection tests of relaxed profiles of high density
clouds in a low density ambient medium.  The resolution is 512 zones
for all runs.  We define $n_{c}\equiv\lambda_{F}/ \delta x$, i.e. the
number of zones in a Field length at the mean density, giving a
measure of the resolution at scales for which conduction is important.
We vary $\mathcal{K}$ so that $n_c$ varies from 8 to 32 in powers of
2.  The initial conditions consist of a top-hat function of high and
low density set to be in approximate pressure equilibrium.  The exact
equilibrium state is the solution of
$\rho\mathcal{L}=\nabla\cdot(\mathcal{K}\nabla T)$ from equation
(\ref{energy}), and is different for each value of ${\cal K}$.
Initial oscillations in pressure, velocity, and density gradually
decay, and we consider the profile to be relaxed when these
oscillations reach approximately one percent of their value early in
the simulation.  After a relaxed ``cloud'' profile is achieved the
velocity of all cells is set to a constant value comparable to the
sound speed, and the ``cloud'' is advected through the grid twice.
Profiles at the end of these runs are compared to the initial relaxed
equilibrium profile in Figure~\ref{fig2}.  Notice that as $n_{c}$
increases so does the number of zones over which the contact
discontinuity is spread; the results clearly show that the profile is
preserved more faithfully as $n_{c}$ increases from 8 to 32.  For
comparison we also show results from the original ZEUS-2D code,
without conduction and cooling.

It is clear that running higher levels of conduction at a given
resolution has the advantage of smearing contact discontinuities over
an increasing number of zones, thus improving the performance of the
code in the advection tests.  However, increasing $\mathcal{K}$ has
the disadvantage of inhibiting thermal instability at larger and
larger spatial scales, such that only very large scale structures can
develop from thermal instability.  As we are interested in how
wavelengths of growing MRI modes in a cloudy medium may be affected by
the distances between condensations, it is undesirable to limit the
available dynamic range for this exploration.  For $n_c$ = 32 at
a resolution of 256 zones, $\lambda_{F}$ is 12.5pc, and the
maximum TI growth rate occurs at 29 pc, about one third of our box.  A
second disadvantage of increasing $\mathcal{K}$ is that conduction
quickly begins to set the time step for the simulations.  So, we make
the practical choice of setting $n_c=8$.  At resolution of $256^2$,
and a box size of 100 pc, we then set $\mathcal{K}=1.03\times 10^{7} \
\rm{erg \ cm^{-1} \ K^{-1} \ s^{-1}}$, to yield $\lambda_F=3.125$
pc. This compromise choice allows us to resolve TI developing from the
initial conditions, to maintain advection profiles within $\sim 20\%$,
and to have adequate numbers and resolution of the condensations that
form to represent a cloudy medium in a meaningful way.

The true thermal conduction level in the ISM is not well known
observationally, and must be affected by the fractional ionization
(since electrons are highly mobile when present) and magnetic field
geometry.  A minimum level of conduction for the atomic ISM is that of
neutral hydrogen gas, $\mathcal{K}=2.5 \times 10^3 \ T^{1/2} \ 
\rm{erg \ cm^{-1} \ K^{-1} \ s^{-1}}$ \citep{par53}.  At 2000 K,
$\mathcal{K}=1.1 \times 10^5 \ \rm{erg \ cm^{-1} \ K^{-1} \ s^{-1}}$, about a
factor of one hundred smaller than our value, such that $\lambda_F$
would be reduced by a factor $\sim 10$.  The use of a smaller
conduction parameter would not, however, significantly alter our main
results.  The most unstable wavelength for TI (see Figure~\ref{fig1})
would be significantly smaller, 3 pc compared to 12 pc for our adopted
$\mathcal{K}$ -- which would reduce the size scale of the clouds in
the initial condensation phase.  However, the growth rate of TI for
the most unstable wavelength would increase by only 15\% compared to
our models.  The evolution towards more massive clouds via
agglomeration would proceed similarly to the results we have found.
In addition, the overall tendency to maintain a two-phase medium after
TI has developed, as well as the characteristics we identify for MRI
growth in a cloudy medium, would not be affected by the initial sizes
of clouds that form.

\section{Thermal Instability Simulations}
\label{TI}
\subsection{Physical Principles and Timescales}

Various heating and cooling processes in the ISM define a thermal
equilibrium pressure-density curve on which energy is radiated at the
same rate it is absorbed. Perturbations from the equilibrium curve 
will either be stable or unstable depending on its local shape
\citep{fie69}.  Our adopted equilibrium curve \citep{vaz00} is shown
in Figure~\ref{fig3} (together with contours of temperature
corresponding to transitions in the cooling function, and with scatter
plots from our first simulation).  Gas in the region above the curve
has net cooling ($\mathcal{L} > 0$), and gas below the curve net
heating ($\mathcal{L} < 0$).  When gas on the equilibrium curve in the
warm phase (phase ``F'', at T $>$ 6102 K) is perturbed to higher
(lower) temperatures, there is net cooling (heating) and the gas
returns to equilibrium.  The same situation applies to gas in the cold
phase (phase ``H'', at T $<$ 141 K).  At intermediate temperatures
(phase ``G'', 313 K $<$ T $<$ 6102 K), however, perturbations from
equilibrium to higher (lower) temperatures results in net heating
(cooling), and the gas continues heating (cooling) until it reaches
equilibrium in Phase F (H).  This is the physical basis for TI, which
was first analyzed comprehensively by \citet{fie65}.

Thermal instability has long been believed to play an important role
in structuring the ISM because at typical volume-averaged atomic
densities estimated for the ISM (e.g. $\rm{n_{HI}} = 0.57$ from
\citet{dic90}) gas in thermal equilibrium would lie on the unstable
portion of the curve.  Thus, much of the mass of the ISM in the Milky
Way and similar galaxies is believed to be in cold clouds or a warm
intercloud medium, the two stable neutral atomic phases
(e.g. \citet{wol03}).  Recent theoretical work has emphasized that
dynamical processes may drive gas away from these two stable phases,
potentially explaining observationally-inferred temperatures that 
depart from equilibrium expectations; we shall discuss these 
issues in \S~\ref{discuss}.
It is, nevertheless, of significant interest to study in detail how TI
develops nonlinearly to establish a two-phase cloudy medium in the
absence of other potential effects such as localized heating, impacts
from large-scale shocks, or stresses associated with MHD turbulence.
Results of these carefully-controlled ``pure TI'' simulations are valuable
for characterizing the timescale to develop a two-phase medium as well
as its structural and kinetic properties.  ``Pure TI'' models also
represent a baseline for comparison of models incorporating more
complex physics.

Once the particular form of the cooling curve has been chosen, the
development of TI is primarily a function of four parameters: the
cooling time, $t_{cool}$, the heating time, $t_{heat}$, the sound
crossing time, $t_{sound}$, and the conduction length scale,
$\lambda_F$.  The cooling time depends on the specific cooling
function; for the cooling curve we adopt \citep{san02}, in varying
temperature regimes we have
\begin{equation}
\label{eq1}
t_{cool}=\frac{\mathcal{E}}{\rho^2\Lambda} = \left\{ \begin{array}{l}
        {1.2\times 10^4 \ \yr} \
        \frac{{(P/k)/2000}}{(n/20)^{2}(T/100 \ \K)^{2.12}} \ \ \ \
        10 \ \K < T  < 141 \ \K\; \\ {1.5\times 10^4 \ \yr} \
        \frac{{(P/k)/2000}}{(n/10)^{2}(T/200 \ \K)^{1.00}} \ \ \ \
        141 \ \K < {T} < 313 \ \K\; \\ {3.7\times 10^5 \ \yr} \
        \frac{{(P/k)/2000}}{(n/1)^{2}(T/2000 \ \K)^{0.56}} \ \ \ \
        313 \ \K < {T} < 6102 \ \K \; \\ {1.1\times 10^6 \ \yr} \
        \frac{{(P/k)/2000}}{(n/0.25)^{2}(T/8000 \ \K)^{3.67}} \ \ \
        \ 6102 \ \K < {T} < 10^5 \ \K \; \\
\end{array} \right. 
\end{equation}

The heating time,
\begin{equation}
t_{heat}=\frac{\mathcal{E}}{\rho\Gamma}=4.1\times 10^5 \ \yr
\frac{{(P/k)/2000}}{(n/1)}
\end{equation}
for $\Gamma =0.015\ \erg\ \s^{-1} \g^{-1}$.
We can also define an effective cooling/heating time as
$t_{cool,eff}=|t_{cool}^{-1}-t_{heat}^{-1}|^{-1}$; at thermal
equilibria, $t_{cool,eff}$ becomes infinite.  The sound crossing time
over distance $\ell$ is
\begin{equation}
t_{sound}=\frac{\ell}{c_s} = 1.1 \times 10^6\ {\rm yr}
\frac{\ell/5\pc}{(T/2000 \ \K)^{1/2}}
\end{equation}

\begin{figure}
\plotone{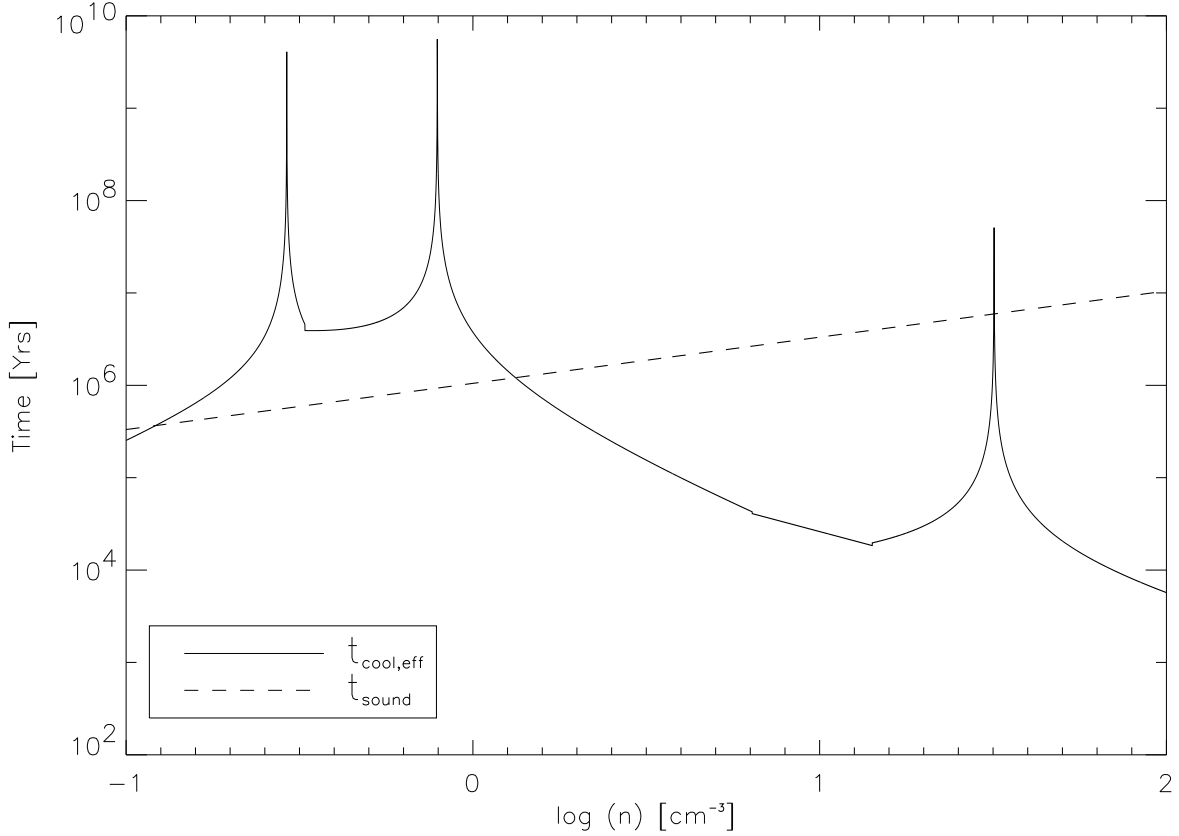}
\caption{Comparison of $t_{cool,eff}$ and $t_{sound}$ as a function of
  density for $P/k=2000$ and $\ell=5\ \pc$.  For range of $0.1 \ \rm{cm^{-3}}
  \lesssim n \lesssim 1 \ \rm{cm^{-3}}$ the sound crossing time is
  less than the effective cooling time, so that gas can evolve at
  approximately constant pressure.  For higher densities,
  $t_{cool,eff}$ is typically an order of magnitude shorter than
  $t_{sound}$, and the gas tends to cool towards the equilibrium
  curve.  The peaks where $t_{cool,eff}$ approaches infinity represent
  the equilibrium densities (n= 0.29 $\rm{cm^{-3}}$, 0.78
  $\rm{cm^{-3}}$, 32 $\rm{cm^{-3}}$) at which heating and cooling
  rates are equal for $P/k=2000$.
\label{times}}
\end{figure}

In Figure~\ref{times}, we plot $t_{cool,eff}$, and $t_{sound}$ for
$\ell=5 \pc$ as a function of density for $P/k=2000\ \K\ \cm^{-3}$.  For
a range of densities $0.1 \ \rm{cm^{-3}} < n < 1 \ \rm{cm^{-3}}$, the
sound crossing time is significantly shorter than the effective
cooling time; otherwise $t_{sound}$ is typically more than an order of
magnitude longer than $t_{cool,eff}$.

To simulate thermal instability under conditions representative of the
general diffuse ISM, we adopt an initial density of $n =1\ \cm^{-3}$
and set the initial pressure to $P/k=2000 \ \K \ \cm^{-3}$; all
velocities are initially set to zero. Random pressure perturbations
with an amplitude less than 0.1\% are added to seed the TI .  The gas
is initially in a weakly cooling state, since the equilibrium pressure
corresponding to $n=1\ \cm^{-3}$ is $P/k=1660 \ \K \ \cm^{-3}$.  The
simulation proceeds for 470 Myr, corresponding to
two galactocentric orbits at the solar circle.

\subsection{Structural Evolution}

The initial development of TI proceeds quickly.  In Figure~\ref{fig3}
we show four snapshots at different times of the density distribution
alongside scatter plots of pressure versus density overlayed on the
equilibrium cooling curve.  Structure begins to form at about $5 \
\rm{Myr}$, and density contrasts continue to increase at constant
pressure until $14 \ \rm{Myr}$ as shown in Figure~\ref{fig3}.  The
Fourier transform of the density distribution at this time,
Figure~\ref{den_pow}, shows that the majority of the power is
concentrated at $k=14$. Allowing for geometrical factors, this is
consistent with the one-dimensional linear theory prediction that the
most unstable wavelength is $k \sim 9$ for our chosen value of
conduction.  A network of filaments briefly forms at about 18 Myr
connecting the regions of highest density, with flow moving along the
filaments increasing the number density to as high as $18 \
\rm{cm}^{-3}$.  By 23 Myr the dense gas has collected in condensations
and has relaxed to approximate thermal equilibrium.  The size scale of
the cold clumps is initially only a few parsecs and is relatively
uniform.  However, random motions of the newly formed clouds leads to
merging and disassociation; conductive evaporation at its boundaries
can also alter the shape of a clump.  By the end of the simulation, at
474 Myr, the typical cloud size has increased significantly to $\sim10
\ \rm{pc}$, though a number of smaller clouds either remain from the
initial TI development or have formed as a result of disassociation.

In Figure~\ref{figxa}, Panels A, B, and C, we plot the volume-weighted
and mass-weighted density probability distribution functions (PDFs)
for the first three snapshots from Figure~\ref{fig3}.  At t=14 Myr, in
Panel A, all of the gas is in the unstable range.  Already, though, in
Panel B, after 18 Myr, the distribution has begun to separate into two
distinct phases.  At t=18 Myr, 69\%, 28\%, and 3\%  of the
gas by volume is in the F (warm), G (unstable), and H (cold) phases
respectively, which are defined by our cooling curve as the ranges 
$n< 0.5 \ \cm^{-3}$, $0.5 \ \cm^{-3}< n < 5.8 \ \cm^{-3}$,
and $5.8 \ \cm^{-3} < n$, respectively.  By mass these proportions
are reversed to 29\%, 23\%, and 49\%, respectively.  From about 100 Myr through
the end of the simulation, the distribution of cloud sizes evolves,
but the PDF remains relatively unchanged, with 12\%, 2\%, and 86\% 
of the mass residing in the F, G, and H phases.

We also performed the same simulation, but increased the resolution to
$512^2$.  The conduction coefficient is not changed, so $n_c=16$.  We
find similar results overall.  In particular, the mass-weighted
density PDFs are compared in Panel D of Figure~\ref{figxa}.  These
PDFs exhibit no significant differences, confirming the robustness of
our results.

\subsection{Thermal Evolution}

Alongside the images of density in Figure~\ref{fig3} we show scatter
plots of $n$ against $P$ for all zones in the grid at the same times.
In the initial state, pressure is constant, and $t_{sound}$ at 5 pc
($\sim 1/2$ the length of the fastest-growing TI mode) is shorter than
$t_{cool}$ (see Figure~\ref{times}).  Towards the low density warm
phase, $t_{sound} << t_{cool}$, and gas parcels in regions undergoing
rarefaction are able to heat nearly isobarically.  Thus, all zones at
densities lower than the mean are filled with an intercloud medium
that maintains spatially nearly uniform pressure.  For gas parcels
undergoing compression and net cooling, as $\delta\rho$ becomes large,
$t_{cool} << t_{sound}$, so the gas tends to cool towards the thermal
equilibrium curve at a faster rate than the flow is able to readjust
dynamically.  After gas parcels reach near thermal equilibrium in the
cold phase, they continue to be compressed until pressure equilibrium
with the warm medium is re-established.  Over time, the average
pressure in the simulation box decreases due to radiation from the
cold phase.

Because the scatter plots in Figure~\ref{fig3} contain a large number
of points, many fall in the unstable range, although the actual amount
of material there is small.  To quantify this, in Figure~\ref{figxb} we
plot mass-weighted and volume-weighted temperature PDFs at times
corresponding to the snapshots in Figure~\ref{fig3}.  As with the
density PDF, the distribution is separates into two distinct phases in
Panel B at 18 Myr, and remains so for the duration of the simulation.

\subsection{Kinetic Evolution}

Thermal instability is a dynamic process, and a number of recent works
have proposed that TI may help contribute to exciting turbulence in
the ISM. In Figure~\ref{fig3a} we plot the mass-weighted
velocity dispersions for gas in the F, G, and H phases separately.
For all phases, the largest velocities occur during the condensation
stage at early times (10-20 Myr), corresponding to about 5 times the
$e$-folding time of the dominant linear instability.  The peak velocity
is about 0.45 $\rm km \ s^{-1}$ for the unstable (G) phase, and is
$\sim 0.3 \ \rm km \ s^{-1}$ for the two stable phases.  At later times
the velocity dispersion in each phase remains relatively constant,
with the largest value (0.35 $\rm km \ s^{-1}$) for phase G, next
largest (0.25 $\rm km \ s^{-1} $) for the warm phase, F, and smallest
(0.15 $\rm km \  s^{-1}$) for the cold phase, H. The standard deviation
of the total velocity dispersion is typically 0.15 $\rm{km \ s^{-1}}$.
For comparison, typical sound speeds of the F, G, and H phases are 7-8
$\rm km \ s^{-1} $ , 1-5 $\rm km \ s^{-1} $ and 0.6-1 $\rm km \
s^{-1}$, respectively.  Thus, the mean turbulent velocities are all
subsonic.

\section{MRI Simulations} 
\label{MRI}
\subsection{MRI Physics}

In a series of four papers, Balbus \& Hawley presented the first linear
analysis and numerical simulations of MRI in the context of an
astrophysical disk  \citep{bal91,haw91,haw92,bal92}.  The physical
basis for the instability is relatively simple, and there are two
requirements for the instability to be present: a differentially
rotating system with decreasing angular velocity as one moves outward
through the disk, and a weak magnetic field (strong magnetic fields
have a stabilizing effect).  As fluid elements are displaced outward
(inward) the magnetic field resists shear and tries to keep the fluid
moving at its original velocity.  Due to these magnetic stresses,
fluid elements gain (lose) angular momentum, the centrifugal force
becomes too large (small) to maintain equilibrium at the new position,
and the fluid element moves farther outward (inward).  This leads to
the transport of angular momentum outward through the disk.

For a complete linear analysis of the MRI in 2D we refer the reader to
\citet{bal91}.  Here we simply summarize the important formulae for
axisymmetric modes with wavenumber ${\bf k}=k_z \hat z$ and
${\bf B}_0=B_0\hat z$.  The growth rates are given by
\begin{equation}
\frac{\gamma^2}{\Omega^2}=\nu^2 \left[\frac{2q-\nu^2}{\nu^2+2-q+[4\nu^2+(2-q)^2]^{1/2})}
\right]
\end{equation}
where $\nu\equiv k_z v_{{\rm A}z} / \Omega$ in terms of the Alfv\'{e}n speed 
$v_{{\rm A}z}\equiv B_{0z}(4\pi\rho)^{-1/2}$. 
The maximum growth rate occurs when
\begin{equation}
\left(\frac{kv_{{\rm A}z}}{\Omega}\right)_{peak}=\frac{(4-(2-q)^2)^{1/2}}{2},
\end{equation}
i.e. $\lambda_{peak}=4\pi v_{{\rm A}z}/(\sqrt{3}\Omega)$ for $q=1$; here
the growth rate is $\gamma_{peak}=\Omega q/2 \rightarrow
\Omega/2$. The highest wavenumber for which axisymmetric MRI exists
when ${\bf B}_0=B_0\hat{z}$ is
\begin{equation}
\left(\frac{kv_{{\rm A}z}}{\Omega}\right)_{max}=\sqrt{2q}.
\label{maxkmri}
\end{equation}

We have tested the code without cooling and conduction, and found that
it can accurately reproduce the predicted linear growth rates of the
MRI.  We do not detail the results here; instead we refer the reader to
\cite{haw92} for a complete analysis of similar models.

Based on the linear dispersion relation, for sufficiently weak
magnetic fields, modes with a range of $k_{z}$ (and also $k_r$) may
grow.  The smallest permissible wavenumber for a simulation is
$(k_z)_{min}=2\pi/L_z$, where $L_z$ is the vertical dimension of the
computational box.  
At late times in 2D axisymmetric simulations \citep{haw92}, MRI
becomes dominated by a ``channel'' solution corresponding to the
smallest permissible vertical wavenumber, i.e. with flow moving
towards the inner regions of the disk on one (vertical) half of the
grid, and flow moving towards the outer regions in the other
(vertical) half of the grid. This pure ``channel flow'' is unphysical;
for a 3D system it is subject to nonaxisymmetric parasitic
instabilities \citep{goo94}. In 3D non-axisymmetric simulations
(e.g. \citet{haw95a}) the channel solution forms at early times, but
later develops into a fully turbulent flow.

The MRI has primarily been studied in the context of accretion disks,
but can be important in any differentially rotating disk system
provided the magnetic fields are not too strong.  In addition to
axisymmetric modes, nonaxisymmetric MRI modes can also grow directly
(see e.g. \citet{bal92} and \cite{kim00} eqs. 80, 81 for instantaneous
growth rates and instability threshold criteria in various limits).
The axisymmetric mode with wavelength $\lambda_z \sim \rm{2H}$ is
the most difficult to stabilize as $B_z$ increases; 
from equation (\ref{maxkmri}), taking
$\Omega = 26 \ \rm{km \ s^{-1} \ kpc^{-1}}$, $q=1$, a disk scale height
$H$=150 pc and a uniform density $n=0.6 \ \cm^{-3}$, MRI will be
present provided $B_z < 0.6 \ \mu G$. For the Milky Way, this is
consistent with the observed solar-neighborhood estimate $|B_z| =
0.37 \ \mu\rm G$ \citep{han99}.
If a multi-phase system behaves similarly to the corresponding
uniform-density medium, then we may expect MRI to be important in the
galactic ISM.  Here, we explore how MRI development can be affected by
strong non-uniformity in the density structure.

\subsection{Evolutionary Development: TI + MRI Model}

To study nonlinear development of the MRI in a nonuniform medium, we first
perform a simulation identical to the TI model run described in
\S~\ref{TI}, but now include magnetic fields and sheared rotation.
All hydrodynamical variables are initialized as described in
\S~\ref{TI}. The magnetic field is vertical with
$\beta=P_{gas}/P_{mag}=1000$. The rotation rate is set to 26 km
$\rm{s^{-1} \ kpc^{-1}}$ representative of the local value near the 
Sun, and we set the shear parameter $q=1.0$ to describe a flat
rotation curve. With these parameters, from equation (\ref{maxkmri}),
the smallest-scale uniform-density MRI mode that would fit within our
$L_z=100 \ \pc$ box has $k_z=3$ (in units of $2\pi/L_z$).

On the left in Figure~\ref{fig3sf} we show snapshots of number density
overlayed with magnetic field lines at three representative times, and
on the right we show the corresponding mass-weighted density PDF.  The
time-scale for development of the MRI is much longer than that of TI,
so that the initial development is essentially the same as in the
purely hydrodynamical case.  During the TI condensation phase the
magnetic field becomes kinked as the filaments condense into small
clouds.  The remaining random motions of the clouds leads to further
distortion of the magnetic field, as can been seen at 237 Myr.  The
channel solution has clearly taken hold by 474 Myr, and the $k_z=1$ mode
(in units of $2\pi/L_z$) dominates.

Similarly to our analysis of kinetic evolution for the TI model, in
Figure~\ref{fig3sb} we plot the velocity dispersion for the F, G, and
H phases as a function of time in the TI + MRI model.  Initially these
are similar to the hydrodynamical case, with all velocities less than
$0.5 \ \rm km \ s^{-1}$.  As the channel solution develops the
velocity dispersion begins to increase at about 500 Myr, and peaks at
the end of the simulation in phase F at approximately $1.5 \ \rm km \ 
s^{-1}$.  The peak velocity dispersion is about $1.2 \ \rm km \ 
s^{-1}$ for phase G, and 0.9 $\rm km \ s^{-1}$ (approximately the
sound speed) for phase H, all towards the end of the simulation.

As in the hydrodynamical model, the PDFs for the TI + MRI model are
clearly two-phase, with small amounts of gas contained in the unstable
regime.  At very late times the fully developed channel solution tends
to increase the proportion of unstable gas.  In
Figure~\ref{mri_temper} we plot the mass-weighted temperature PDF of
the TI + MRI model 800 Myr, and the same quantity for the TI run at
474 Myr.  We do not expect that the PDF for the TI run would evolve
significantly if the simulation had been continued to 800
Myr.  Evidently, the dynamical flows induced by the MRI can
significantly affect the temperature distribution. The larger velocity
dispersion and kinked magnetic fields due to the channel solution can
compress portions of cold clouds, decreasing the temperature
correspondingly.  Although still dominated by distinct warm and cold
phases, there is also a higher proportion of gas in the unstable
regime.  For the same model snapshot, Figure~\ref{mri_scatter} shows a
$P/k$ vs. $n$ scatter plot, overlayed on the equilibrium cooling curve.
In the high density regime, cooling times are short, and the gas is not
far from equilibrium.  At low densities cooling times are longer, and
gas can be found out of thermal equilibrium.

Since the 2D channel solution would break up in a real 3D disk due to
parasitic instabilities \citep{goo94}, we do not expect the late-time
effects seen in our models to have direct implications for the
temperature distribution in the diffuse ISM.  They illustrate, however,
the more generic point that spatially-varying rotational shear coupled
to magnetic fields can create stresses that force gas away from the
stable equilibrium phases.  We shall discuss this further,
highlighting differences that might be expected for 3D MRI, in
\S~\ref{discuss}.

Important questions for assessing MRI development in a cloudy medium
are how the spatial- and time-scales of the fastest-growing modes
differ compared to those in single-phase counterpart systems.  We
measure MRI mode amplitudes in the simulations by taking the Fourier
transform of $B_y$ as a function of time, from which we can calculate
the growth rates.  The $k=1, 2$ and 3 mode amplitudes are plotted in
Figure~\ref{fig_ti_mhd_mode}.  There is not an obvious linear stage
from which we can measure the growth rate, but between 284 and 470 Myr
the average $k=1$ growth rate is about 0.28 $\Omega$, compared to the
predicted rate of 0.34 $\Omega$ at the average density of the model.
At the average density linear theory predicts that the most unstable
mode is at $k=2$, with a growth rate of 0.50 $\Omega$, and the $k=3$
mode is unstable as well, with a predicted growth rate of 0.41
$\Omega$. Between 284 and 470 Myr we measure a mean growth rate for
the $k=2$ and $k=3$ modes to be 0.12 $\Omega$ and 0.18 $\Omega$,
respectively.  At the density of the warm medium, only the $k=1$ mode
is predicted to be unstable in our simulations, with a growth rate of
0.45 $\Omega$.  Thus, growth rates of available modes appear somewhat
lower than they would be for a medium at either the mean density or
the density of the warm medium. In addition, initial growth does not
show the clear dominance of a single fastest growing mode that is
evident in comparison adiabatic test simulations for a single-phase
medium.  However, at late times, the $k=1$ mode grows to exceed the
other low-order modes, similarly to the findings of \citet{haw92} for
a single-phase medium.

\subsection{Evolutionary Development: Cloud + MRI Model}

Because the initial MRI growth in the previous model may be strongly
affected by lingering dynamical effects of TI, it is of interest to
consider MRI development in a medium which contains two distinct
phases from the outset. In our next simulation we therefore begin with
a two phase medium in approximate equilibrium, rather than developing
a two phase medium from thermally unstable gas.  We embed 59 high
density clouds in a low density ambient medium such that the average
density is the same as that of previous TI simulations.  To set up the
initial conditions allowing for conduction at cloud/intercloud
interfaces, we first create ``template'' cloud profiles by embedding a
single high density cold cloud in a low density warm medium and
evolving until a thermally- and dynamically-relaxed state is reached.
This profile (density, pressure and velocity) is then copied to
randomly chosen locations on the grid, with the condition that cloud
centers must be at least 20 zones apart.  We initialize the magnetic
field after this ``cloud embedding'' procedure.  The simulation is
then evolved as the MRI develops.

In Figure~\ref{gmc_ave} we show three snapshots from the ``cloud +
MRI'' simulation, along with mass-weighted density PDFs.  The MRI
takes about 500 Myr until its development begins to become apparent,
as can be seen in the poloidal field lines at 464 Myr in
Figure~\ref{gmc_ave}.  At 701 Myr many of the clouds have merged and
have significant velocities as the MRI channel solution begins to take
hold.  Initially the mass weighted density PDF shows almost no gas in
the unstable range, but as the MRI begins to develop, this phase
begins to become populated, and the PDF becomes very similar to those
from the TI + MRI simulations.

Perhaps the most interesting results from this ``cloud + MRI''
simulation are the behavior of the mode amplitudes and growth rates.
Initially the clouds contain negligible velocities, and are in what
would be a steady state if magnetic fields were not present.  We might
expect, then, to find ``cleaner'' MRI growth rates than for the TI +
MRI runs.  The mode amplitudes for $k=1, 2$, and  3 are plotted in
Figure~\ref{gmc_ave_mode}.  Initially the $k=2$ mode is dominant and shows
approximate linear growth between 300 Myr - 700 Myr, with an
average growth rate of 0.34 $\Omega$, compared to the predicted rate
0.50 $\Omega$ at the average density.  The $k=1$ mode also shows
approximate linear growth with an average rate of 0.47 $\Omega$
measured from 450 Myr - 900 Myr, and becomes the dominant mode at
around 800 Myr.  At the density of the warm medium the theoretical
growth rate of the $k=1$ mode is 0.45 $\Omega$, and at the
average density it is 0.34 $\Omega$.  The $k=3$ growth rate, measured
between 400 Myr and 800 Myr is 0.33 $\Omega$, which we can compare to
the predicted value at the average density of 0.41 $\Omega$.  Thus,
similarly to the TI + MRI model, growth rates are slightly lower than
they would be in a medium with the same mean density.  There is,
however, a longer period of dominance by the mode with the fastest
expected growth rate.

For comparison we also performed an MRI simulation with cooling and
conduction disabled, initially at uniform density and seeded with the
velocity and pressure profiles from the initial state of the previous
``cloud + MRI'' simulation; we also performed zero-conduction,
zero-cooling test simulations seeded with random perturbations. The
growth rates are in reasonable agreement between these simulations,
although the case directly seeded with random perturbations had
cleaner growth of mode amplitudes. Most importantly, we find that
rather than $k=2$ predominating as we found in the ``cloud + MRI''
simulation, the $k=3$ mode in the single-phase comparison model is
dominant until late time.  This suggests that the initial
perturbations are strongest at $k=3$, but the presence of the
multi-phase medium in the ``cloud + MRI'' model significantly inhibits
the growth of this mode because the inertial load varies strongly over
a wavelength for $k=3$.

\subsection{Perspective: Effects of Cloudy Structure}

There are a number of ways MRI growth rates and preferred scales could
be affected by the influence of a cloudy background density structure,
and signs of these effects are evident in our simulations. First, both
growth rates and preferred scales are dependent on the Alfv\'{e}n
speed, which is a function of density.  The cold, dense clouds will be
MRI unstable at small scales compared to the larger preferred scales
of the warm, low density, ambient medium.  For fixed magnetic field
strength, MRI wavelengths are inversely proportional to density.
Thus, MRI wavelengths in cold, dense gas will be a few parsecs, while
MRI wavelengths in warm, diffuse gas will be several tens of parsecs.
Although the MRI wavelengths in dense gas may be smaller than
individual cold clouds, permitting initial rapid growth, further
development of the small-scale instability is limited by clouds' small
radial extent.  Long term MRI development must therefore have
characteristic wavelengths representative of either the average
density conditions or the pervasive low-density warm, intercloud gas.
This expectation is indeed consistent with our results.  As seen in
Figures~\ref{fig3sf} and \ref{gmc_ave}, both the diffuse gas and the
cold clouds participate together in an overall large-scale flow.  The
cold clouds frequently are the sites of strong kinks in the magnetic
field.

One might also expect the preferred scales for MRI to be affected by
cloud spacing.  If cloud spacing is small compared to a given
wavelength, then the MRI growth rate might be expected to be similar
to that under average density conditions.  Thus, if the
fastest-growing wavelength at the mean density is large compared to
cloud spacing along a field line then this might be expected to be the
dominant wavelength.  This situation is indeed evident in the second
frame (464 Myr) of Figure~\ref{gmc_ave}, in which the $k=2$ mode
dominates, even though, as described above, the input perturbation
spectrum is such that the $k=3$ mode would dominate if the density
were uniform.  On the other hand, if fewer, more massive clouds exist,
cloud separations will be larger, which could suppress MRI growth at
wavelengths shorter than cloud spacings and encourage MRI development
on the largest scales.  Evidence of this effect can be seen in the
third panel (474 Myr) of Figure~\ref{fig3sf}.

Finally, if total mass is distributed very unevenly with magnetic
flux, then MRI may develop more rapidly and at longer wavelengths in
regions where there is a comparatively low inertial load.  In
simulations (not shown) we have performed which have alternating
radial zones of high and low mass loading on field lines (initiated
with cold clouds at intersections in a Cartesian grid), we indeed see
this effect.  Development of MRI in the low-inertia ``pure warm''
phase is, however, checked when the radially-moving flow collides with
the high-inertia cold clouds.

To test whether the growth rate of the smaller-scale k=2 mode
(essentially near the lower wavelength limit for behavior as at a
single average density) could be enhanced in TI+MRI models, 
we also performed an
additional random TI simulations in which a $k=2$ perturbation was added
shortly after the initial condensation phase.  The perturbation was
added at about the 20 percent level in $v_{1}$.  For the first 500
Myr the $k=2$ mode is strongest with a growth rate of 0.36
$\Omega$.  After 500 Myr the k=1 mode, with a growth rate of
0.47 $\Omega$, is again dominant.

Taken together, the simulations of this section show that the
development of the MRI in the presence of a cloudy medium has 
modest differences compared to the corresponding development in a
single phase medium at the average density.  The dominant wavelengths
are similar to those predicted by linear theory at mean-density
conditions, and growth rates are also similar, but slightly smaller.
The spacings between clouds affects which among the low-$k$ modes 
dominates the power during the exponential-growth phase.
At very late times the $k_z=1$ mode is dominant in all simulations,
consistent with the ultimate dominance of this channel flow in the
single-phase models of \citet{haw92}.

\section{Summary and Discussion} 
\label{discuss}

Thermal and magnetorotational instabilities may play a major role in
determining the physical properties of the diffuse ISM.  In regions
far from active star formation or a recent supernova explosion, TI and
MRI may even be the primary processes driving structure and dynamics
in the ISM on scales $\simlt 100$ pc.  The PDFs of gas density and
temperature, the characteristic sizes, shapes, and spatial
distributions of cloudy structures, and the amplitudes and spectral
properties of turbulent velocities and magnetic fields may all be
strongly influenced by TI and MRI.  In addition, development and
saturation of TI and MRI may be strongly interdependent.  In this
paper, we have initiated a study of these important processes using
numerical MHD simulations.  The current work focuses on code tests and
2D models using a microphysics implementation appropriate for the
atomic ISM.  In addition to characterizing the properties of TI and
MRI modes in their nonlinear stages, this study lays the groundwork
for future 3D simulations which will be used to investigate
quasi-steady turbulence.

In the following, we summarize the results presented herein, 
compare to other recent work, and discuss key issues for 
future investigation.

1. {\it Numerical methods:} 
We have implemented atomic-ISM heating/cooling and
thermal-conduction source terms in the energy equation of the ZEUS
code (using implicit and explicit updates, respectively).  For
conditions representing the mean pressure and density in the ISM, we
find excellent numerical agreement with the analytic growth rates of
thermally-unstable modes for a large range of wavelengths and thermal
conductivity coefficients.  Based on these tests and confirmation of
acceptable results for advection of high-contrast contact
discontinuities (warm/cold pressure equilibrium interfaces) on the
grid, we adopt a value of ${\cal K}=10^7\ {\rm erg}\ \cm^{-1}\
\s^{-1} \ \K^{-1}$ such that the Field length is resolved by 8 (16) zones in
(100 pc)$^2$ simulations with 256$^2$ (512$^2$) cells.

Explicit inclusion of conduction is important for suppressing
numerically-unresolved TI-driven amplification of grid-scale noise;
\citet{koy03} have also recently highlighted the importance of
implementing conduction for simulations of thermally bistable media.
In some previous simulations of TI \citep{kri02a,kri02b} under stongly
cooling conditions, conduction was not included; since those
simulations began with relatively large-amplitude (5\%) perturbations
on resolved scales, however, sub-dominant effects from unresolved
growth at grid scales in the initial stages of TI would be less
noticeable.  In other recent work \citep{vaz03} simulations of TI
using spectral algorithms (with explicit diffusive terms in the
equations of motions) appear to have difficulty reproducing the
analytic growth rates in some circumstances.  Conceivably, this
may be a sign of numerical diffusion that could tend to produce more
gas in thermally-unstable regimes than is realistic, in simulations
using these computational methods.

2. {\it Nonlinear development of TI:} 
In ``pure TI'' simulations where we initialize gas at $P/k=2000\
\K\ \cm^{-3}$ and $n=1\ \cm^{-3}$ in a (100 pc)$^2$ box with 0.1\%
initial pressure perturbations, we find that TI develops at a
characteristic length scale consistent with the predicted
fastest-growing mode, $\sim 12$ pc for our adopted value of $\cal K$.
As seen in other 2D simulations (e.g. \citet{vaz00}), the structure
initially resembles a ``honeycomb'' network of cells, and as nonlinear
development proceeds, gas condenses into cold, compact clouds at the
intersections of filaments.  Gas undergoing rarefaction towards the
warm phase heats nearly isobarically, because the sound crossing time
is short compared to the net heating-cooling time.  Gas undergoing
compression towards the cold phase initially has isobaric evolution
(while density perturbations remain low-amplitude), but then tends
first to cool toward the equilibrium curve very rapidly (with an
attendant pressure drop), and then dynamically readjusts its density
and temperature until the pressure again matches ambient conditions.
The time to establish a distinct two-phase structure of well-separated
cold clouds within a warm ambient medium (see third panel of Fig. 4)
is $\sim 20$ Myr, or about 10 $e$-folding times in terms of the linear
growth rate.  In the subsequent evolution, the cold, dense clouds
undergo successive mergers to produce larger structures.

The transition from nearly isobaric to more ``isochoric''-like
evolution for cold gas during nonlinear stages of condensation was
recently emphasized by \citet{bur00}, and snapshots of phase diagrams
in \citet{kri02a} show a similar dip in pressure for overdense gas as
it cools toward thermal equilibrium.  \citet{vaz03} found, similar to
our results, that initial perturbations of similar or larger sizes to
our dominant TI wavelength require times $>10$ Myr to complete the
condensation process, even when a much larger (10\%) initial
perturbations are used.  The real level of conduction in the atomic
ISM may be lower than the value we adopted (for numerical efficacy),
with the fastest-growing TI wavelength a factor $\sim 4$ smaller than
our 12 pc value and the condensation time correspondingly shorter;
\citet{san02} found that 3 pc-scale overdensities condense into clouds
within 4 Myr.  As the turbulent cascade is likely to maintain
nonlinear-amplitude entropy perturbations down to sub-pc scales, we
expect that the fastest-growing wavelength \footnote{From
\citet{fie65}, this is essentially the geometric mean of $\lambda_{\rm
F}$ and the product of the sound speed and the cooling time.}  is
likely to dominate when TI occurs under ``natural'' circumstances,
with later mergers producing larger clouds (see also \citet{san02}).

3. {\it Gas phase distributions from TI:} The bimodal density and
temperature PDFs in our TI simulations mirror the distinct two-phase
structure evident in late-time snapshots.  Typical late-time warm-,
cold-, and intermediate-temperature mass fractions are 12, 86, and
2\%.  For a two-phase medium with mean density $\bar n$ and cold and
warm densities $n_{\rm c}$ and $n_{\rm w}$, the fraction of mass in
the cold medium is $f_{\rm c}=(1-n_{\rm w}/\bar n)(1- n_{\rm w}/n_{\rm
c})^{-1}$.  Provided $n_{\rm c} \gg n_{\rm w}$, the mass fraction in
the warm medium is thus $f_{\rm w}\approx n_{\rm w}/\bar n$.  Since
the pressure at late stages of our evolution has dropped near the
minimum value of $P$ at which two phases are present, and in thermal
equilibrium at this pressure $n_{\rm w}\approx 0.1 \ \cm^{-3}$ (with $n_{\rm
c}\approx 10 \ \cm^{-3}$), the relative proportions of gas $f_w\approx 0.1,
f_c\approx 0.9$ in the cold and warm phases are just as expected (with
$\bar n =1 \ \cm^{-3}$).

Findings on density and temperature PDFs from other recent TI
simulations are varied.  From the 3D simulations of \citet{kri02a,
kri02b}, the late-time (1.5 Myr) mass fractions are $f_{\rm w}=0.42$,
$f_{\rm c}=0.44$ in the stable phases and $f_{\rm i}=0.14$ in the
intermediate, unstable regime.  Kritsuk and Norman use a somewhat
different cooling curve from ours, with $n_{\rm w}=0.4 \ \cm^{-3}$ in
thermal equilibrium at the minimum pressure at which two stable phases
are available.  Since they use $\bar n=1 \ \cm^{-3}$, their result that
$f_{\rm w}\approx n_{\rm w}/\bar n$ is consistent with expectations
for a two-phase medium, while the gas at intermediate temperatures
appears to be due to mass exchange with the cold medium (see their
discussion).

In the 1D simulations of \citet{san02} (and using the same cooling
curve and mean density as ours), only a few percent of the gas in
their ``multiple condensation'' runs remains at intermediate
densities, similar to our results, and their $f_{\rm w}=0.3-0.4$ at
$t\sim 20-25$ Myr is similar to our results at comparable (early)
times.  In the 2D simulations of \citet{gaz01} that also include
``stellar-like'' local heat sources, the late-time mass fractions are
$f_{\rm w}=0.25$, $f_{\rm c}=0.25$ $f_{\rm i}=0.50$.  It is not clear
to what extent this large proportion of gas at intermediate
temperatures is sustained by turbulence (via adiabatic
expansion/compression and/or shocks heating or cooling gas that would
otherwise be in the warm or cold stable phases), versus being
maintained by the localized heating turned on when $n>15 \cm^{-3}$.
\footnote{Since real star formation is confined to giant molecular
clouds rather than occurring in a more distributed fashion in cold
atomic clouds, localized stellar heating (and turbulent driving by
expanding HII regions) may have much less impact on HI density and
temperature PDFs in the real ISM.}  With 3D MRI simulations in which
turbulent driving is ``cold'', it will be possible to address this
important issue.
 
4. {\it Turbulent driving by TI:} We find that turbulence produced by
``pure TI'' has only modest amplitudes, when initiated from ``average
ISM'' pressure and density conditions.  For the warm, unstable, and
cold phases, respectively, we find typical mass-weighted velocity
dispersions of $0.25$, $0.35$, and $0.15$ $\kms$.  These velocities
are all quite subsonic.  In simulations starting from thermal
equilibrium, \citet{kri02a} similarly find subsonic turbulence (${\cal
M}_{rms}\sim 0.3$ at $t<$ 2Myr), although when gas is initially very
hot, supersonic turbulence can be produced.  When they include
repeated episodes of strong UV heating \citet{kri02b} find Mach number
variations ${\cal M}_{rms}\sim 0.2-0.6$ between ``low'' and ``high''
states; since their ``low state'' is dominated by cold gas with
$c_s\sim 1 \ \kms$, this is consistent with our results for typical
turbulent amplitudes.  In the simulations of \citet{koy02} in which
warm gas shocks on impact with a low-density, hot ($T=3\times 10^5$ K)
layer, TI develops near the interface of shocked gas with the hot
medium, leading to the formation of cold cloudlets with velocity
dispersions of a few $\kms$.  Although Koyama and Inutsuka attribute
this turbulence to the effects of TI, it is possible that other
dynamical instabilities associated with the hot/warm interface
contribute in driving these motions.

5. {\it Nonlinear development of axisymmetric MRI:} 
We have studied the development of axisymmetric MRI under atomic ISM
conditions, both with ``TI+MRI'' models starting from uniform density
and pressure ($P/k=2000\ K\ \cm^{-3}$ and $n=1\ \cm^{-3}$), and with
``cloud+MRI'' models that are initiated with the same uniform pressure
and total mass, but start with a population of cold clouds embedded in
a warm ambient medium.  The magnetic field in both types of models is
vertical and initially uniform, with $B^2/8\pi=P/1000$.  The peak
growth rate of MRI (in a uniform medium) is $\Omega/2$, where $\Omega$
is the local angular velocity of the galaxy.  Since this growth rate
is a factor $\sim 40$ lower than typical TI growth rates, the early
development of the TI+MRI model is the same as in the ``pure TI''
model.  By the time MRI begins to develop (after a few $100$ Myr), the
TI+MRI model has similar cloud/intercloud structure -- except with
more variations in cloud size -- to the cloud+MRI model.  At early
times, the density and temperature PDFs are essentially the same as
those produced by TI; at late times, however, while the PDFs remain
bimodal, the dense gas is distributed over a somewhat larger range of
densities and temperatures, due to the dynamics of the ``channel
flow'' solution (see below).

6. {\it Spatial scales of MRI in a cloudy medium:} 
In both our TI+MRI and cloud+MRI simulations, after a few
galactic orbital times, the velocity and magnetic fields become
dominated by large-scale structures.  Since the smallest-scale MRI
mode that would fit in our $L_z=100\pc$ box under {\it uniform-density}
conditions has vertical wavenumber $k=3$ (in units $2\pi/L_z$; i.e.
wavelength $\lambda=L_z/3$), and the fastest-growing mode would have
$k=2$, this implies, consistent with expectations, that cloudy density
structure in the supporting medium does not grossly alter the
character of MRI.  We quantify MRI structural development in terms of
mode amplitudes of $B_y$, the azimuthal magnetic field.  For the
TI+MRI model, the amplitudes of the $k=1,2,$ and $3$ modes are all
similar -- and motions in the $x-z$ plane continue to be dominated by TI
effects, with cloud agglomeration -- until $t\sim 400 \ \Myr$, after
which the clouds have become highly concentrated and the $k=1$ MRI
mode associated with the ``channel solution'' \citep{haw92} takes hold.  
For the cloud+MRI model, on the other
hand, the $k=2$ mode grows first (with clouds remaining small and
distributed) and it dominates until $\sim 800 \ \Myr$, when the channel
solution ($k=1$) begins to take precedence.

These differences show that the spatial distribution of clouds can
have a significant effect on selecting which MRI modes are important.
If intercloud distances are small compared to its wavelength, the
dominant MRI mode is the same as that predicted for uniform-density
conditions.  If, however, other turbulent processes acting on scales
{\it small} compared to MRI wavelengths (and times small compared to
$\Omega^{-1}$) collect the clouds and correspondingly increase their
separations, then only MRI modes at scales larger than twice the
typical intercloud distance will be able to grow.  As a consequence,
for MRI to play an important role in the ISM, either the majority of
the gas must remain in a warm, diffuse phase, or else if it collects
in clouds their separations must not be too large.

It is interesting to relate these constraints to observational
inferences of the HI spatial distribution. From the \citet{hei03} HI
absorption observations that yielded 142 separate cold gas components
on 47 lines of sight at $|b|>10^\circ$, their mean separation would be
$\sim 40$ pc (taking the cold disk semi-thickness $\sim 100$ pc).  The
distribution of warm gas is much harder to interpret, but in the
limiting situation where it is mainly in overdense clouds\footnote{According to Heiles and Troland, of the 60\% of the HI
that is in warm gas, $>50\%$ at high latitudes is at lower
temperatures than the $T\sim 8000\K$ required for approximate pressure
equilibrium with the cold clouds; since significant underpressures are
difficult to achieve, this gas is likely to be in clouds denser than
$\bar n$.}, and
using Heiles and Troland's finding that $\sim 25\%$ of emission
components have no associated absorption, the mean distance between
clouds would be $\sim 30$ pc.  Intercloud separations similar to these
estimates are small enough that vertical MRI modes could be supported;
if cloud spacings are appreciably larger, however, they could not be.

7. {\it Growth rates and saturation amplitudes of MRI:} For the
low-$k$ modes that are present in both our TI+MRI and cloud+MRI
models, typical growth rates are generally comparable to those for
modes of the same wavelength in a medium of the same mean density.
For the TI+MRI model, typical growth rates are measured to be $0.28 \ 
\Omega$, $0.12 \ \Omega$, and $0.18 \ \Omega$ for the $k=1$, 2, and 3
modes, respectively, compared to the rates $\gamma/\Omega=0.45$,
$0.5$, and $0.41$ that would apply for a uniform medium.  For the
cloud+MRI model, the exponential MRI growth is ``cleaner''; rates are
$\gamma/\Omega=0.47$, $0.34$, and $0.33$ for $k=1$, 2, and 3 modes,
respectively.  The growth rates of smaller-scale ($k=2,3$) modes are
thus slightly more affected by the presence of cloudy structure than
that of the largest-scale ($k=1$) mode, consistent with expectations.

Although definitive results await 3D simulations, these findings
provide support for the possibility that MRI may drive turbulence in
the diffuse ISM at amplitudes consistent with observations of HI
emission and absorption.  From previous 3D simulations under
relatively {\it uniform} conditions (accomplished by adopting an
isothermal equation of state), the velocity dispersions driven by MRI
in steady-state were found to be smaller than observed values.  In
particular, \citet{kim03} found that the typical 1D turbulent
amplitudes are 3 - 4 $\kms$, whereas the observed nonthermal
contribution to the 1D velocity dispersion for both cold and warm gas
amounts to $\sigma_v\sim 7\ \kms$ \citep{hei03}.  Thus, for a {\it
single phase medium}, MRI-driven turbulent velocity amplitudes in
steady state -- which are determined by a balance between excitation
and dissipation -- fall a factor $\sim 2$ short of explaining
observations.

Since our present cloudy-medium models show growth rates quite
comparable to those in a one-phase medium, the key question is
therefore whether MRI dissipation rates are reduced in a cloudy
medium, and if so, whether the reduction can yield a factor two
increase in $\sigma_v$.  To see that a quantitative effect at this
level is not unreasonable, consider the comparison to an idealized
system of ${\cal N}_{cl}\equiv \ell^{-3}$ clouds per unit volume having
individual radii $r$, internal density relative to the mean value
$n_{cl}/\bar n$, and RMS relative velocity dispersion $\sigma_v$.
With turbulent energy driving and dissipation rates $\dot{\cal
E}_{in}$ and $\sim \sigma_v^2/t_{coll}$, where the collision time 
$t_{coll}= (4 \sqrt{\pi} r^2 {\cal N}_{cl} \sigma_v)^{-1} $, 
$\sigma_v$ in steady state is
an order-unity factor times 
$(\dot {\cal E}_{in} \ell)^{1/3}(n_{cl}/\bar n)^{2/9}$.  For this 
idealized situation,
concentrating material into clouds with $n_{cl}/\bar n \sim 30$
(similar to cold ISM clouds)
would indeed increase $\sigma_v$ by
a factor two compared to the case with near-uniform conditions, 
$n_{cl}/\bar n \sim 1$.  With 3D
simulations, it will be possible to test whether a similar scaling
behavior holds for the saturated state of 
MRI-driven turbulence in cloudy vs. single-phase ISM models.

\acknowledgments

We are grateful to Woong-Tae Kim, Jim Stone, and Mark Wolfire for
valuable discussions.  This work was supported in part
by grants NAG 59167 (NASA) and AST 0205972 (NSF).

\clearpage

\clearpage 

\clearpage 

\begin{figure}
\epsscale{0.82} 
\caption{
Density evolution in the TI simulation. 
{\it  Left}: snapshots of $\log(n)$ at representative times as noted. {\it
    Right}: scatter plots of $n$ and $P/k$ from the simulation, together
  with the equilibrium cooling curve and the labeled temperature
  contours that demark the transitions between the F, G, and H phases.
\label{fig3}}
\end{figure}

\begin{figure}
\epsscale{1.0} 
\plotone{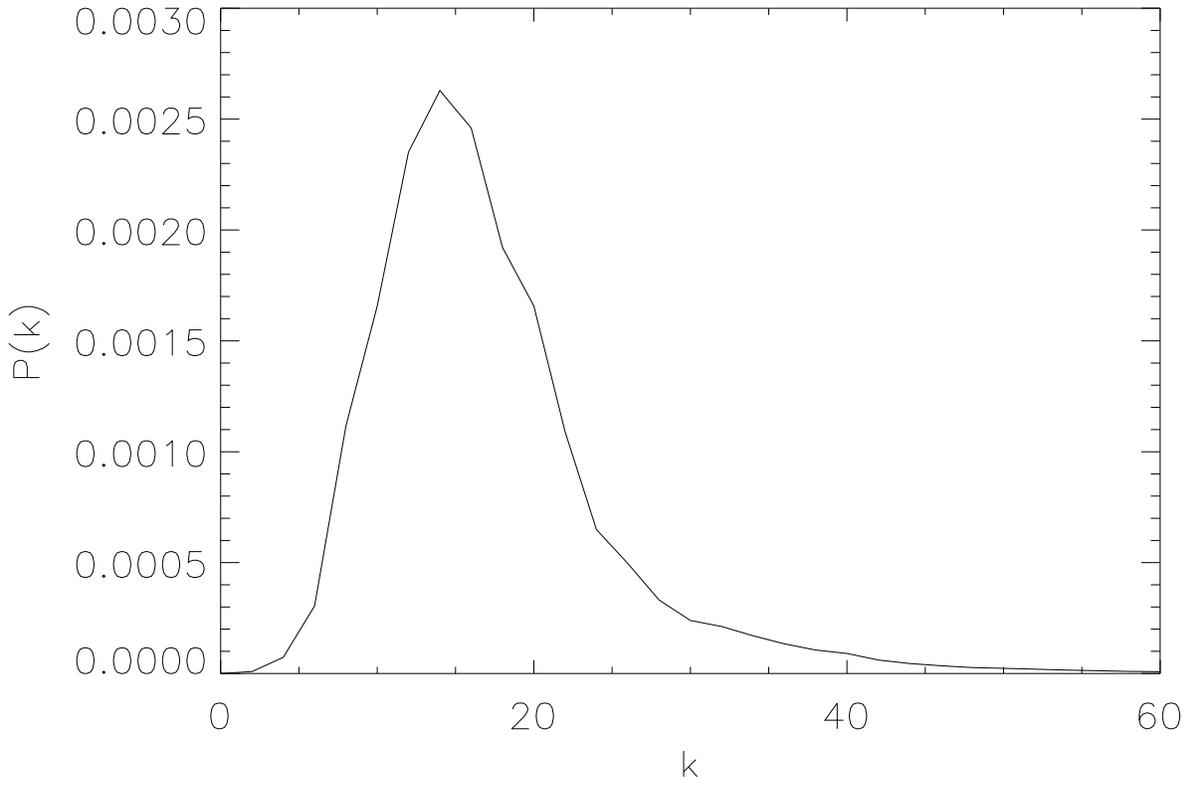}
\caption{
Power spectrum of the TI density distribution at 14 Myr.
\label{den_pow}}
\end{figure}

\begin{figure}
\epsscale{1.0}
\plotone{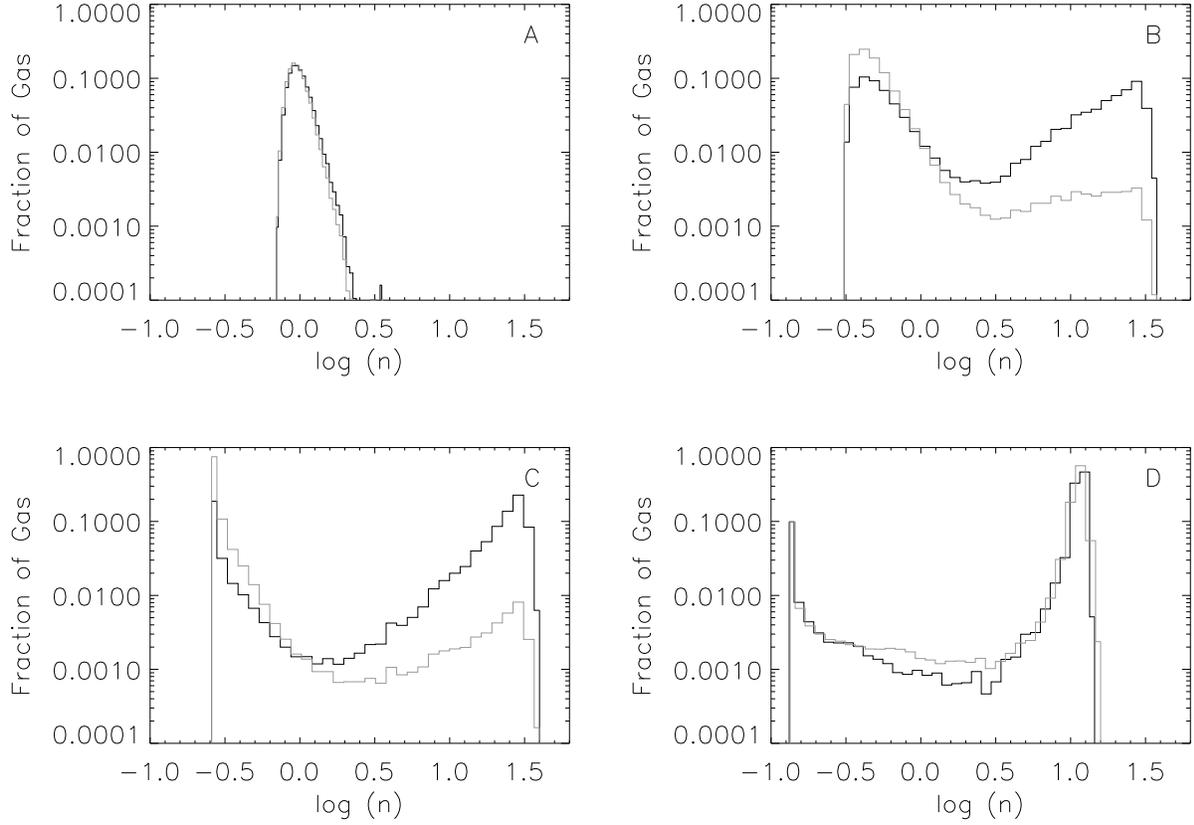}
\caption{Panels A,B and C show the mass (black line) and volume (grey
  line) density PDFs for TI at the same times as the first three
  snapshots in Figure~\ref{fig3}.  Panel D compares the mass weighted
  density PDF for the standard resolution of $256^2$ (black line) and
  $512^2$ (grey line) at time 474 Myr.
\label{figxa}}
\end{figure}

\begin{figure}
\plotone{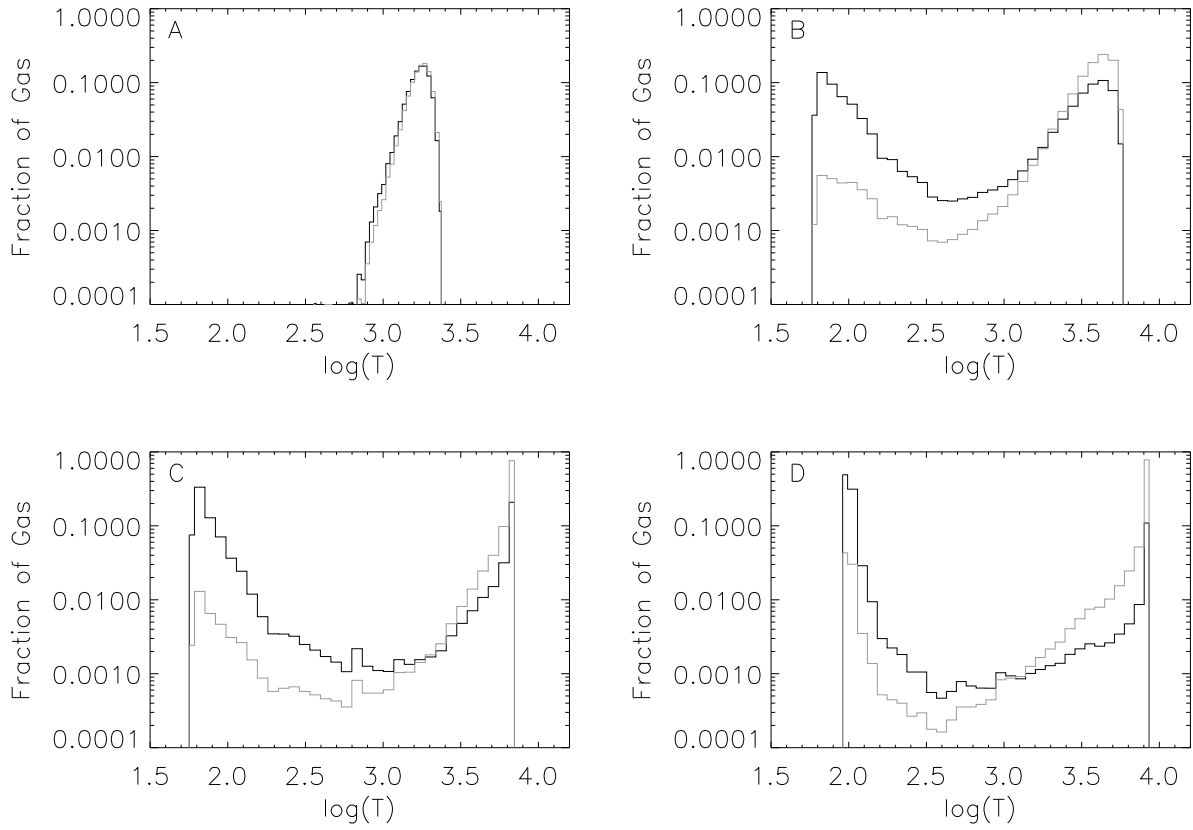}
\caption{Mass (black line) and volume (grey line) temperature PDFs for
  the snapshots in Figure~\ref{fig3} (time increases A-D).
\label{figxb}}
\end{figure}

\begin{figure}
\plotone{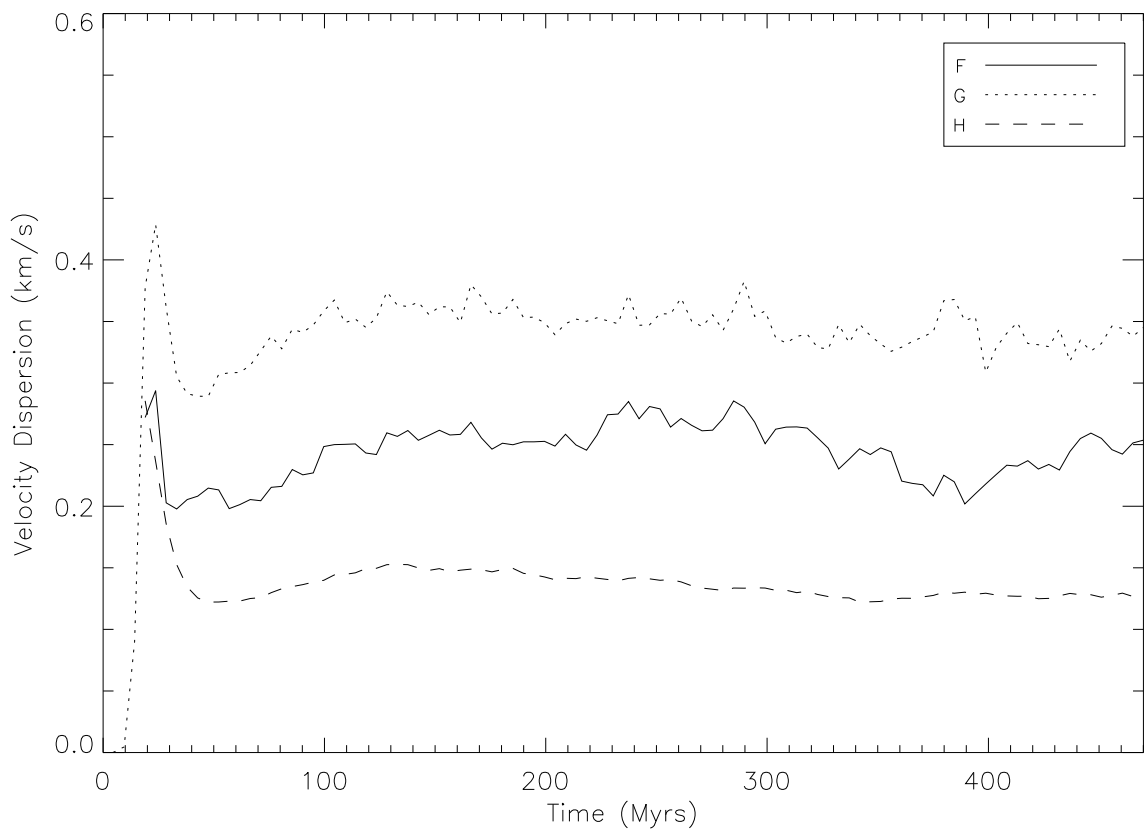}
\caption{Evolution of velocity dispersion = $(v_x^2+v_z^2)^{1/2}$ in
  the TI model.  Typical velocities for the F (warm), G (unstable),
  and H (cold) phases are $0.35 \ \km\ \s^{-1}$, 
  $0.25 \ \km \ \s^{-1}$, and $0.15 \ \km \ \s^{-1}$. 
\label{fig3a}}
\end{figure}

\begin{figure}
\epsscale{0.85}
\caption{Structural evolution of the TI + MRI simulation.  {\it Left}:
  snapshots of $\log(n)$ at representative times as noted, overlayed with
  magnetic field lines. {\it Right}: mass-weighted density PDF at the
  same times as the snapshots.  
\label{fig3sf}}
\end{figure}

\begin{figure}
\epsscale{1.0}
\plotone{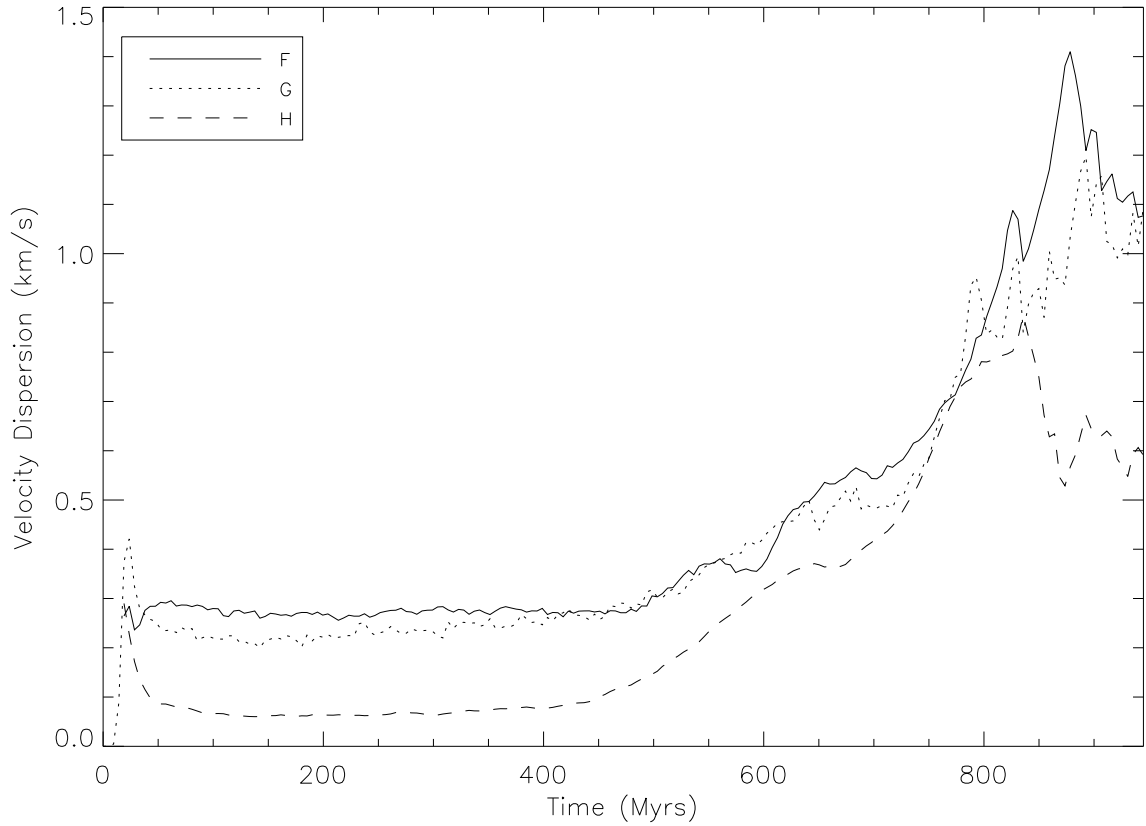}
\caption{Mass-weighted velocity dispersion = $(v_x^2+v_z^2)^{1/2}$ for
  TI + MRI model, separated by phase. The initial velocity dispersion
  is due to the development of TI (see Figure~\ref{fig3a}).  After 400
  Myr the MRI becomes important, and the channel solution increases
  the velocity dispersion to as high as 1.4 $\km \ \s^{-1}$ towards
  the end of the simulation.
\label{fig3sb}}
\end{figure}

\begin{figure}
\epsscale{1.0}
\plotone{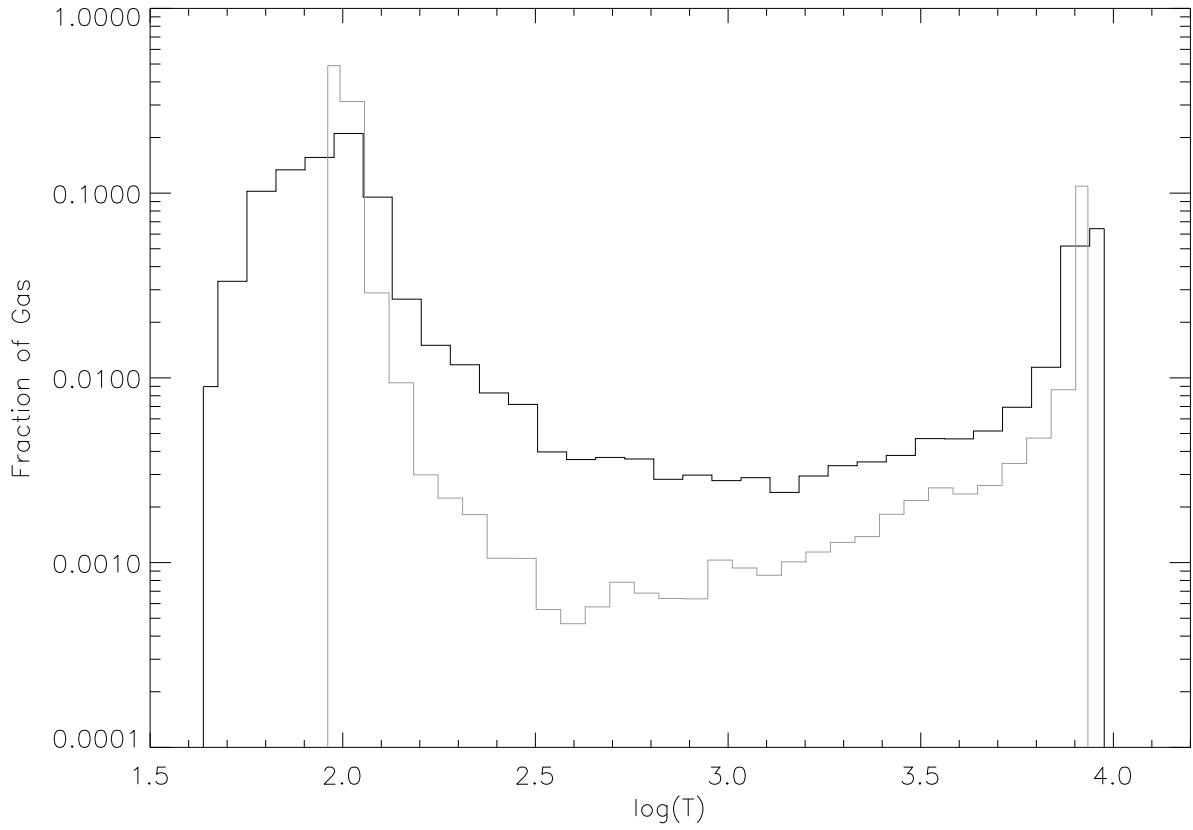}
\caption{Mass-weighted temperature PDF for the TI run at 474 Myr (grey
  line) and TI + MRI at 800 Myr (black line), after the channel
  solution has fully developed.  The active dynamics of MRI leads to
  the presence of high density/low temperature gas.
\label{mri_temper}}
\end{figure}

\begin{figure}
\epsscale{1.0}
\plotone{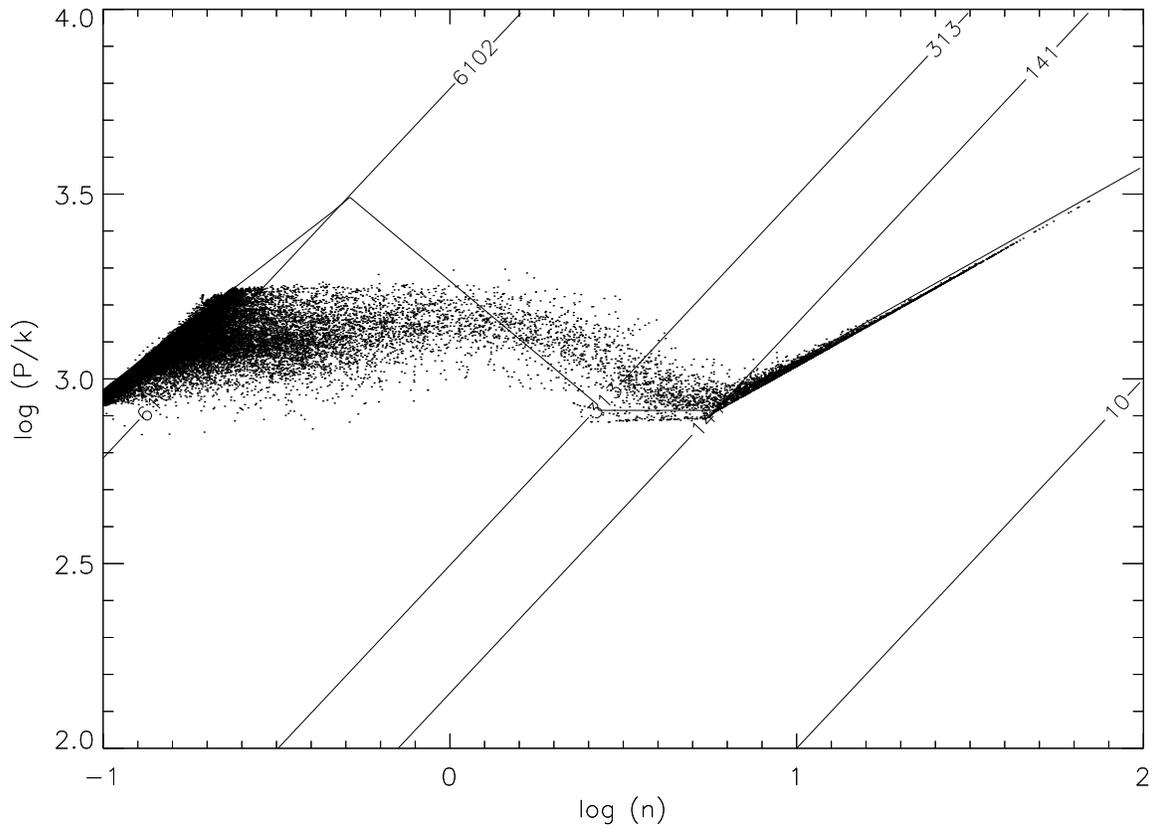}
\caption{Scatter plot of $n$ and $P/k$ at 800 Myr, near the end of the
  simulation, for the TI + MRI model.  Cooling timescales are short
  for the high density gas, so gas remains near thermal equilibrium
  for a range of pressures.  In the low density regime, cooling
  time-scales are longer, and there can be significant departures from
  thermal equilibrium. 
\label{mri_scatter}}
\end{figure}

\begin{figure}
\epsscale{1.0}
\plotone{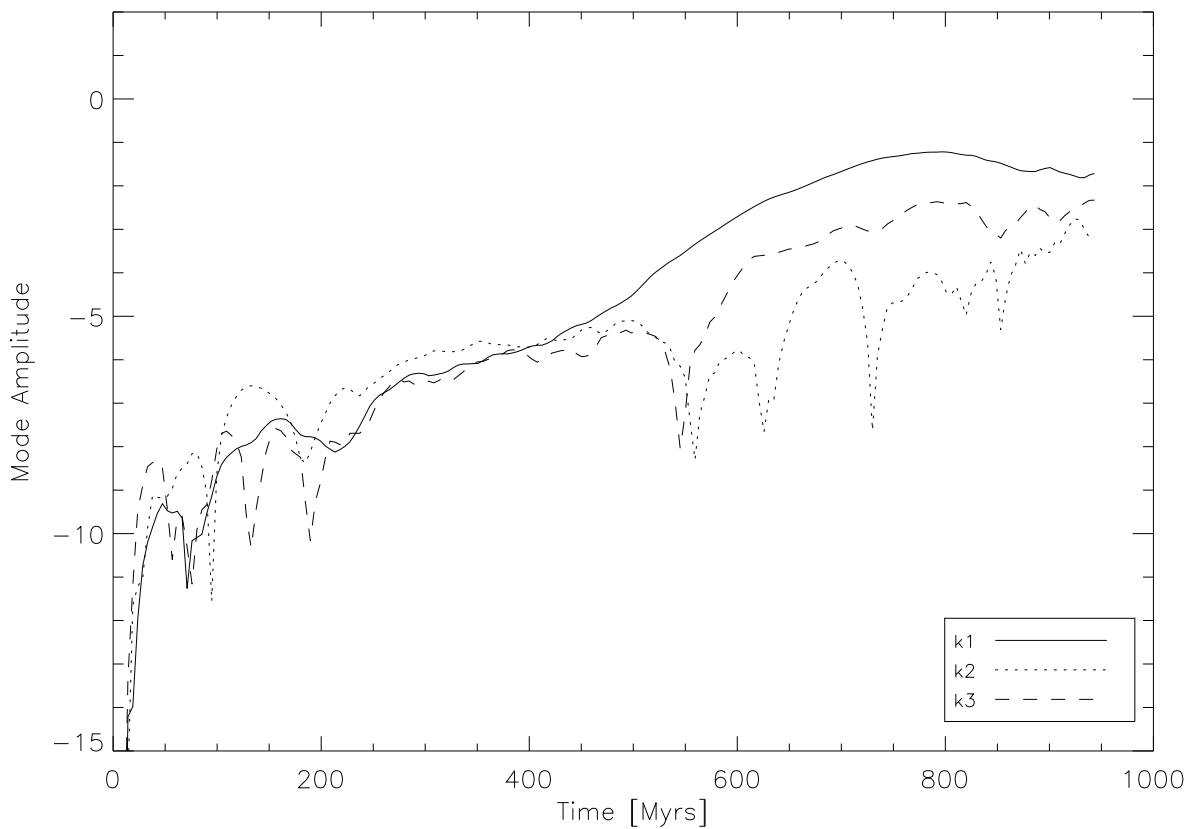}
\caption{Mode amplitudes for $B_y$ in the TI + MRI model.  For the first
  500 Myr the $k=1,2$ and 3 modes are approximately equal in
  amplitude, but the $k=1$ mode is dominant after this point.
\label{fig_ti_mhd_mode}}
\end{figure}

\begin{figure}
\epsscale{0.85}
\caption{Structural evolution of cloud + MRI simulation.  {\it Left}:
  snapshots of $\log(n)$ at representative times as noted, overlayed with
  magnetic field lines. {\it Right}: mass-weighted density PDF at the
  same times as the snapshots.  
\label{gmc_ave}}
\end{figure}

\begin{figure}
\epsscale{1.0} 
\plotone{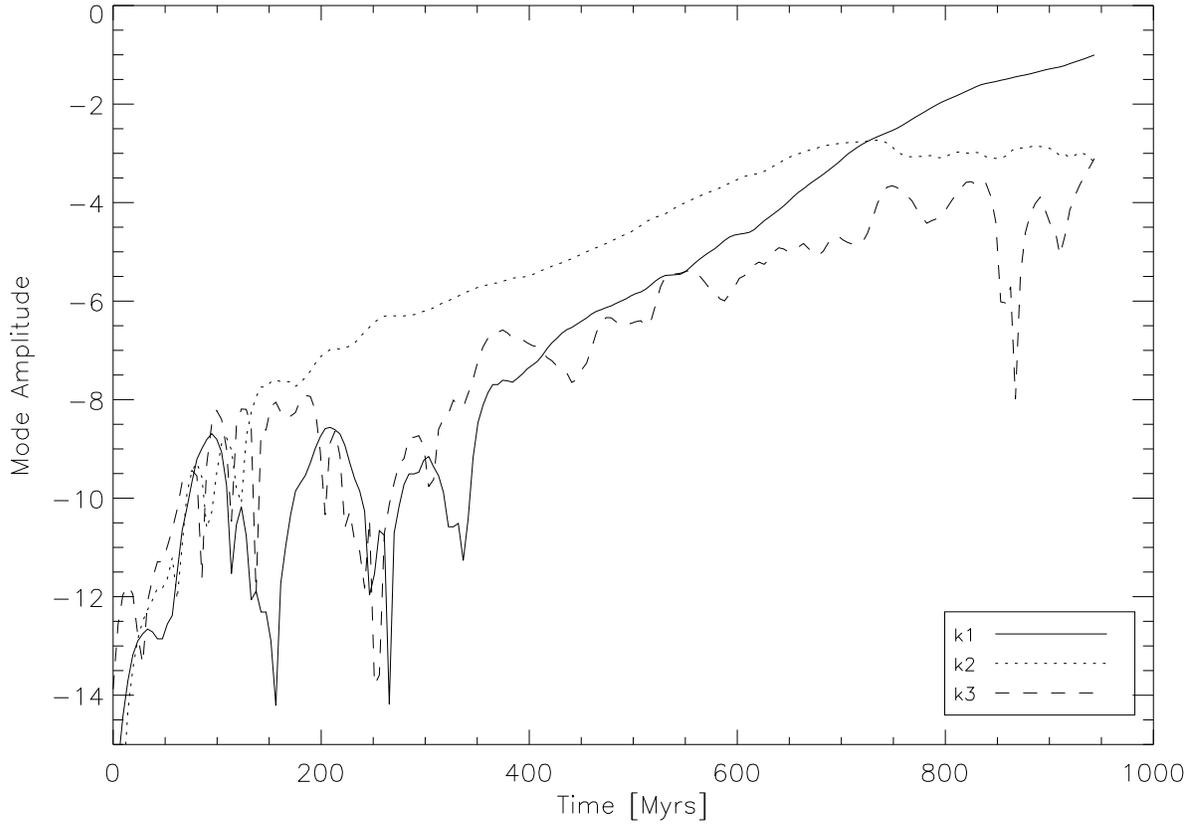} 
\caption{Mode amplitudes of $B_y$ in the cloud + MRI simulation.  The
  $k=2$ mode is initially the largest, but by the end of the simulation
  the $k=1$ mode has become dominant. 
\label{gmc_ave_mode}}
\end{figure} 

\end{document}